\documentclass[final]{siamltex}
\usepackage{epsfig}
\usepackage{graphicx}
\usepackage{amsmath}
\usepackage{amssymb}
\usepackage{setspace}
\usepackage{graphicx}
\graphicspath{{pics/}{figs/}}
\usepackage{graphicx}

\newtheorem{defn}{\noindent $\mathbf{Definition}$}[section]

\newtheorem{thm}[defn]{$\mathbf{Theorem}$}

\title{Beltrami representation and its applications to texture map and video compression}

\author{Lok Ming Lui, Ka Chun Lam, Tsz Wai Wong, Xianfeng Gu}

\begin{document}

\maketitle

\begin{abstract}
Surface parameterizations and registrations are important in computer graphics and imaging, where 1-1 correspondences between meshes are computed. In practice, surface maps are usually represented and stored as 3D coordinates each vertex is mapped to, which often requires lots of storage memory. This causes inconvenience in data transmission and data storage. To tackle this problem, we propose an effective algorithm for compressing surface homeomorphisms using Fourier approximation of the Beltrami representation. The Beltrami representation is a complex-valued function defined on triangular faces of the surface mesh with supreme norm strictly less than 1. Under suitable normalization, there is a 1-1 correspondence between the set of surface homeomorphisms and the set of Beltrami representations. Hence, every bijective surface map is associated with a unique Beltrami representation. Conversely, given a Beltrami representation, the corresponding bijective surface map can be exactly reconstructed using the Linear Beltrami Solver introduced in this paper. Using the Beltrami representation, the surface homeomorphism can be easily compressed by Fourier approximation, without distorting the bijectivity of the map. The storage memory can be effectively reduced, which is useful for many practical problems in computer graphics and imaging. In this paper, we proposed to apply the algorithm to texture map compression and video compression. With our proposed algorithm, the storage requirement for the texture properties of a textured surface can be significantly reduced.  Our algorithm can further be applied to compressing motion vector fields for video compression, which effectively improve the compression ratio.
\end{abstract}


%\begin{AMS}
%15A15, 15A09, 15A23
%\end{AMS}

\pagestyle{myheadings}
\thispagestyle{plain}
\markboth{Lok Ming Lui et al.}{Beltrami representation and applications}

\section{Introduction}
Surface registration and parameterization are important processes in computer graphics and imaging, where 1-1 correspondences between meshes are computed. For example, in video compression, registrations between image frames are necessary to capture the deformation of objects in images \cite{Keller,Fukunaga,Tekalp}. While in computer graphics, surface parameterizations are needed for texture mapping \cite{Levy,Zhang,Haker}. There are many different approaches for surface registration and parameterization. A commonly used method is to find surface maps satisfying certain constraints, such as matching landmarks \cite{Miller,Lui,Lui10,Luilandmark} and minimizing distortions \cite{Gu1,Gu2,Gu3,Hurdal}. Surface maps computed from these processes can be highly convoluted and are usually represented and stored as 3D coordinate functions in $\mathbb{R}^3$. A huge storage memory is therefore required, especially when a large set of fine surface meshes are to be processed. For instance, in computer graphics, mesh parameterization that maps each vertex of the surface to a 2D position on an image is required for texture mapping. In order to have a high-quality textured mesh, mesh parameterizations of high-resolution are necessary. Usually, a great amount of memory and bandwidth are needed to store and transmit the data of the surface map, which causes a great deal of inconvenience. Nevertheless, very little work has been done on the compression of bijective surface maps. This motivates us to look for a compression scheme for surface homeomorphisms, which can significantly reduces the storage requirement.

In this work, we propose an effective algorithm for compressing surface homeomorphism using the Beltrami representation. The Beltrami representation is a complex-valued function defined on triangular faces of the surface mesh with supreme norm strictly less than 1. It measures the local conformality distortion of the surface map. Every surface map is associated with a unique Beltrami representation. According to the Quasi-conformal Teichm\"{u}ller theory, under suitable normalization, there is an 1-1 correspondence between the set of surface homeomorphisms and the set of Beltrami representations. In other words, every surface map can be represented by a unique Beltrami representation. Conversely, given a Beltrami representation, one can reconstruct the unique surface map associated to it. In this paper, we propose an algorithm called the {\it Linear Beltrami Solver} to reconstruct the surface map associated to a given Beltrami representation. Using the Beltrami representation, 1/3 of the required storage space for a bijective surface map is saved. Furthermore, the Beltrami representation has very little constraints. The only constraint is that its supreme norm has to be strictly less than 1. It does not have any requirements of injectivity nor surjectivity. This allows us to further compress the Beltrami representation using Fourier approximations, without distorting the bijectivity of the map. The storage memory can then be significantly reduced. However, Fourier compression is not possible for other representations such as 3D coordinate functions, as the bijective property (1-1 and onto) of the resulting maps cannot be guaranteed (see Figure \ref{fig:Grid_xy_fail}, Figure \ref{fig:BCcoordinatecompare} and Figure \ref{fig:brainfourier}).

Our proposed algorithm for surface map compression can be practically applied to problems in computer graphics and imaging. In this paper, we propose to apply the algorithm to texture map compression and video compression. With our proposed algorithm, the storage requirement for the texture properties of a textured surface can be significantly reduced.  Our algorithm can further be applied to compressing motion vector fields for video compression, which effectively improve the compression ratio.

In short, the contribution of this paper is three-folded: 1. we propose a compression algorithm for surface homeomorphism using the Fourier approximation of the Beltrami representation; 2. we propose the Linear Beltrami Solver to exactly reconstruct a surface map from its associated Beltrami representation; and 3. we apply the proposed algorithm to texture map compression and video compression, which significantly reduces the storage requirement. 

This paper is laid out as follow. In Section 2, we describe the relevant works closely related to this paper. In Section 3, we describe some basic mathematical concepts related to our algorithms. In Section 4, we describe in detail the main algorithm we use to compress bijective surface maps with their Beltrami representations. We also describe how surface maps can be efficiently and accurately reconstructed. Applications of the proposed algorithm to texture map compression and video compression are described in Section 5. In Section 6, a conclusion is drawn.

\section{Related works}
In this section, we give an overview of the previous works mostly related to our paper.

\paragraph{$\mathbf{Surface\ parameterization/registration}$} Surface parameterization and registration have been extensively studied, for which different representations of bijective surface maps have been proposed. Conformal registration, which minimizes angular distortion, have been widely used to obtain a smooth 1-1 correspondence between surfaces \cite{Fischl2, Gu1, Gu3, Haker, Hurdal, Gu2, book}. For example, Hurdal et al. \cite{Hurdal} proposed to compute the conformal parameterizations using circle packing and applied it to registration of human brains. Gu et al. \cite{Gu1, Gu3, Gu2} proposed to compute the conformal parameterizations of human brain surfaces for registration using harmonic energy minimization and holomorphic 1-forms. Conformal registration is advantageous to local geometry preservation. But it cannot match feature landmarks, such as sulcal landmarks on the human brains, consistently. To alleviate this issue, Wang et al. \cite{Wang05,Luilandmark} proposed to compute the optimized conformal parameterizations of brain surfaces by minimizing a compounded energy. All of the above algorithms represent surface maps with their 3D coordinate functions. Special attention is required to ensure the bijectivity of the resulting registration. Besides, smooth vector field has also been proposed to represent surface maps. Lui et al. \cite{Lui10} proposed the use of vector fields to represent surface maps and reconstruct them through integral flow equations. They obtained shape-based landmark matching harmonic maps by looking for the best vector fields minimizing a shape energy. The use of vector fields to represent surface maps makes optimization easier, but they cannot describe all surface maps. Time dependent vector fields can be used to represent the set of all surface maps. For example, Joshi et al. \cite{Joshi} proposed the generation of large deformation diffeomorphisms for landmark point matching, where the registrations are generated as solutions to the transport equation of time dependent vector fields. The time dependent vector fields facilitate the optimization procedure, although they may not be a good representation of surface maps since they requires more storage memory.

\paragraph{$\mathbf{Texture\ mapping}$} The technique of texture mapping has been extensively employed on rendering 3D graphics in animation and video gaming. The basic idea of it is to map an image onto a given surface so as to increase the realism of the 3D model \cite{Texturemappingintro1,Texturemappingintro2}. The problem of finding a suitable parameterization for texturing a polygonal mesh has been widely studied \cite{Bennis,SDMa}. Besides, the storage memory for texture properties contribute a significant portion of the total file size of a textured mesh. Texture mapping compression is therefore necessary. Recently, much attention has been focused on compressing texture images while preserving the quality of the textured 3D model. Balmelli et al. \cite{Balmelli} warped the texture and sub-sampled it to reduce the unnecessary image space. Yoon et al. \cite{Yoon} proposed a wavelet-based image compression technique which is specialised for compressing texture properties of a polygonal mesh. Iourcha et al. \cite{INH99} proposed the S3TC texture compression method, which is the most common texture compression algorithm nowadays. S3TC is a lossy compression algorithm that adapts a block truncation coding. Recently, researches has also been carried out on the high dynamic range (HDR) texture compression, which encodes texture images at a constant 8 bits per pixel \cite{HDR}. Furthermore, to overcome the bandwidth problem in 3D rendering, mesh compression algorithms, which reduce the storage memory for mesh geometry and mesh connectivity, have been proposed \cite{Deering,Taubin}.  However, compression of texture coordinates has received less attention. Isenburg et al. \cite{Isenburg} proposed an algorithm in which texture cooridnates are predicted from mesh vertices through parallelogram rule. However, such algorithm only reduces half of the storage requirement and depends greatly on the similarities between meshes and the texture map.

\paragraph{$\mathbf{Video\ compression}$} Video compression has developed rapidly over last decades. Many techniques and algorithms for video compression have been proposed \cite{Hoang,LeGall,Wangvideo}. The basic idea to achieve the compression is to remove temporal and spatial redundancies existing in video sequences. The first generation of video compression involved techniques for intra-frame coding and simple inter-frame
coding \cite{Gha90}. Motion-compensated predictive coding was later proposed, which has been exploited in all recent video coding standards such as MPEG-2, MPEG-4 or H.264 \cite{Gir87,Hoang,Wangvideo}. The major component in motion compensated coding is the motion vector (MV) estimation. Various techniques for motion vector estimation have been proposed \cite{Koga,Jain,JKuo,Zafar,Zafar2,Zaccarin}. In this paper, we are interested in compressing the motion vectors using their Beltrami representations which can further improve the compression ratio of existing video compression algorithms.

\paragraph{$\mathbf{Surface\ map\ compression}$} 

Compression of mappings has also been studied. Chai et al. \cite{Chai04} proposed the depth map compression algorithm by encoding mappings as a simplified triangular meshes. Lewis \cite{Lewis} described a technique for compressing surface potential mapping data using transform techniques. All these methods deal with the compression of real-valued functions defined on 2D domains. For vector-valued functions, Stachera et al. \cite{Stachera} developed an algorithm to compress normal maps by decomposing them in the frequency domain. Ioup \cite{Ioup} also proposed to compress vector map data in the frequency domain. Kolesnikov et al. \cite{Kolesnikov} proposed an algorithm for distortion-constrained compression of vector maps, based on optimal polygonal approximations and dynamic quantizations of vector data. All these methods do not deal with preserving bijective maps between surfaces. The bijectivity (1-1, onto) of the maps can be easily lost due to lossy compression.

Compression of surface registrations that preserves the bijectivity was preliminary studied by Lui et al. \cite{LuiCompression}. In that work, Beltrami coefficient (BC) defined on every vertex of the mesh was proposed to represent bijective surface maps. BC can approximate the associated surface map well only when the triangulation is regular. Also, Beltrami Holomorphic Flow (BHF) was used to iteratively reconstruct surface maps from their BCs. Integration has to be computed in each iteration, which causes inefficiency in many practical applications. BHF was further applied for the optimization of surface registration in \cite{LuiBHF}. In this paper, we propose to use Beltrami representation defined on each triangular faces to represent bijective surface maps. The Beltrami representation can exactly represent the surface map, regardless of the regularity of the triangulation. Also, Linear Beltrami Solver is proposed to efficiently reconstruct surface map from its Beltrami representation.

\section{Mathematical Background}

In this section, we describe some basic mathematical concepts related to our algorithms. For details, we refer the readers to \cite{Gardiner}\cite{Lehto}\cite{DiffGeomBook}.

A surface $S$ with a conformal structure is called a \emph{Riemann surface}. Given two Riemann surfaces $M$ and $N$, a map $f:M\to N$ is \emph{conformal} if it preserves the surface metric up to a multiplicative factor called the conformal factor. An immediate consequence is that every conformal map preserves angles. With the angle-preserving property, a conformal map effectively preserves the local geometry of the surface structure. %\cite{PDE:JLee}\cite{Add:Chern}\cite{Add:DOCARMO}.

A generalization of conformal maps is the \emph{quasi-conformal} maps, which are orientation preserving homeomorphisms between Riemann surfaces with bounded conformality distortion, in the sense that their first order approximations takes small circles to small ellipses of bounded eccentricity \cite{Gardiner}. Thus, a conformal homeomorphism that maps a small circle to a small circle can also be regarded as quasi-conformal. Surface registrations and parameterizations can be considered as quasi-conformal maps. Mathematically, $f \colon \mathbb{C} \to \mathbb{C}$ is quasi-conformal provided that it satisfies the Beltrami equation:
\begin{equation}\label{beltramieqt}
\frac{\partial f}{\partial \overline{z}} = \mu(z) \frac{\partial f}{\partial z},
\end{equation}
\noindent for some complex valued Lebesgue measurable $\mu$ satisfying $||\mu||_{\infty}< 1$. $\mu$ is called the \emph{Beltrami coefficient}, which is a measure of non-conformality. In particular, the map $f$ is conformal around a small neighborhood of $p$ when $\mu(p) = 0$. Infinitesimally, around a point $p$, $f$ may be expressed with respect to its local parameter as follows:
\begin{equation}
\begin{split}
f(z) & = f(p) + f_{z}(p)z + f_{\overline{z}}(p)\overline{z} \\
& = f(p) + f_{z}(p)(z + \mu(p)\overline{z}).
\end{split}
\end{equation}

\begin{figure*}[t]
\centering
\includegraphics[height=1.60in]{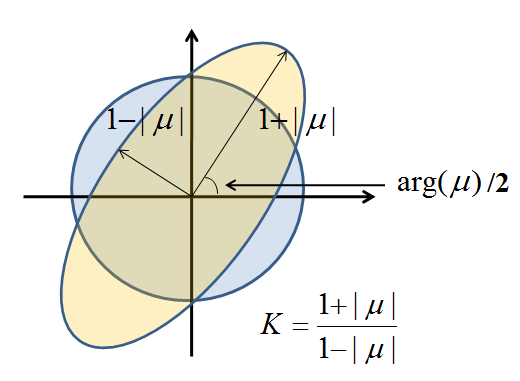}
\caption{Illustration of how the Beltrami coefficient measures the conformality distortion of a quasi-conformal map. \label{beltramicd}}
\end{figure*}

Obviously, $f$ is not conformal if and only if $\mu(p)\neq 0$. Inside the local parameter domain, $f$ may be considered as a map composed of a translation to $f(p)$ together with a stretch map $S(z)=z + \mu(p)\overline{z}$, which is postcomposed by a multiplication of $f_z(p),$ which is conformal. All the conformal distortion of $S(z)$ is caused by $\mu(p)$. $S(z)$ is the map that causes $f$ to map a small circle to a small ellipse. From $\mu(p)$, we can determine the angles of the directions of maximal magnification and shrinking and the amount of them as well. Specifically, the angle of maximal magnification is $\arg(\mu(p))/2$ with magnifying factor $1+|\mu(p)|$; The angle of maximal shrinking is the orthogonal angle $(\arg(\mu(p)) -\pi)/2$ with shrinking factor $1-|\mu(p)|$. The distortion or dilation is given by:
\begin{equation}
K = {\left(1+|\mu(p)|\right)}/{\left(1-|\mu(p)|\right)}.
\end{equation}

Thus, the Beltrami coefficient $\mu$ gives us all the information about the properties of the map (See Figure \ref{beltramicd}).

Given a Beltrami coefficient $\mu:\mathbb{C}\to \mathbb{C}$ with $\|\mu\|_\infty < 1$. There is always a quasiconformal mapping from $\mathbb{C}$ onto itself which satisfies the Beltrami equation in the distribution sense \cite{Gardiner}. More precisely,

\medskip

\begin{thm}[Measurable Riemann Mapping Theorem] \label{thm:Beltrami}
Suppose $\mu: \mathbb{C} \to \mathbb{C}$ is Lebesgue measurable satisfying $\|\mu\|_\infty < 1$, then there is a quasiconformal homeomorphism $\phi$ from $\mathbb{C}$ onto itself, which is in the Sobolev space $W^{1,2}(\mathbb{C})$ and satisfies the Beltrami equation \ref{beltramieqt} in the distribution sense. Furthermore, by fixing 0, 1 and $\infty$, the associated quasiconformal homeomorphism $\phi$ is uniquely determined.
\end{thm}

\medskip

By reflection, the above theorem can be further extended to Beltrami coefficients defined on the unit disk $\mathbb{D}$.

\medskip

\begin{thm}\label{thm:Beltramidisk}
Suppose $\mu: \mathbb{D} \to \mathbb{C}$ is Lebesgue measurable satisfying $\|\mu\|_\infty < 1$, then there is a quasiconformal homeomorphism $\phi$ from the unit disk to itself, which is in the Sobolev space $W^{1,2}(\Omega)$ and satisfies the Beltrami equation \ref{beltramieqt} in the distribution sense. Furthermore, by fixing 0 and 1, the associated quasiconformal homeomorphism $\phi$ is uniquely determined.
\end{thm}

\medskip

Theorem \ref{thm:Beltrami} and Theorem \ref{thm:Beltramidisk} suggest that under suitable normalization, a homeomorphism from $\mathbb{C}$ or $\mathbb{D}$ onto itself can be uniquely determined by its associated Beltrami coefficient. These two theorems can be extended to homeomorphisms between Riemann surfaces \cite{LuiBHF}.

\begin{figure*}[t]
\centering
\includegraphics[height=1.50in]{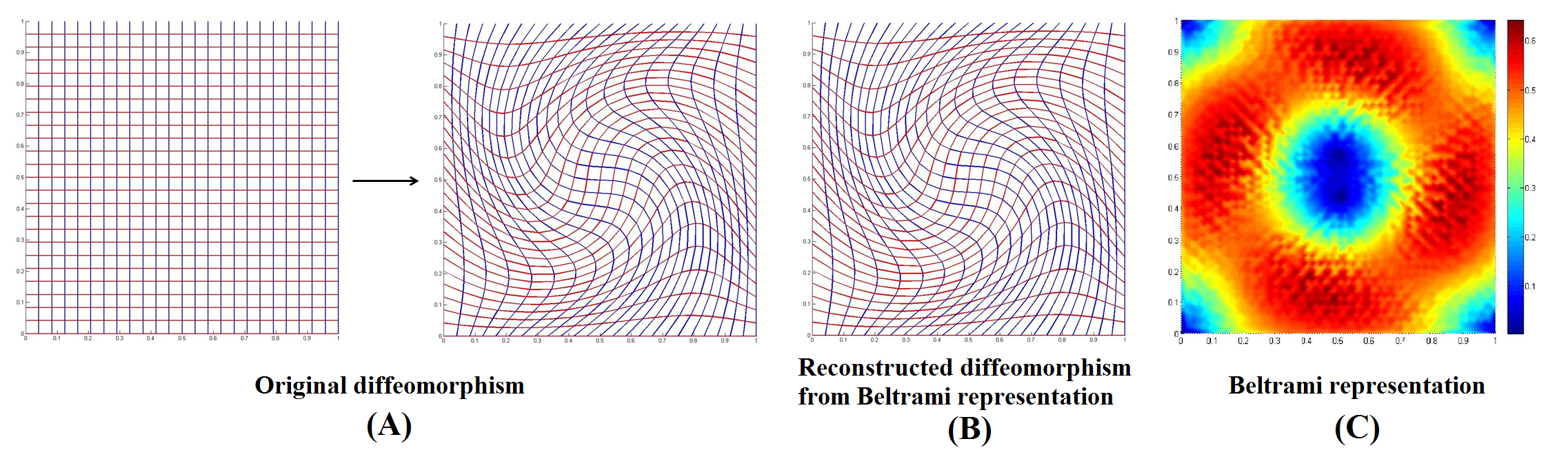}
\caption{Reconstruction of a bijective map from its Beltrami coefficient. (A) shows the original bijective map between the unit square. (B) shows the reconstructed map from its Beltrami coefficient. (C) shows the colormap of the norm of the Beltrami coefficient. \label{fig:xyexample}}
\end{figure*}

Let $S_1$ and $S_2$ be two genus-0 closed surfaces with a 3-point correspondence between them: $\{p_1, p_2, p_3 \in S_1\} \leftrightarrow \{q_1, q_2, q_3 \in S_2\}$. By Riemann mapping theorem, $S_1$ and $S_2$ can both be uniquely parameterized by $\phi_1\colon S_1 \to \mathbb{S}^2\cong\overline{\mathbb{C}}$ and $\phi_2\colon S_2 \to \mathbb{S}^2\cong\overline{\mathbb{C}}$ respectively, such that $\phi_1(p_1)=0,\phi_1(p_2)=1, \phi_1(p_3)=\infty$ and $\phi_2(q_1)=0,\phi_2(q_2)=1, \phi_2(q_3)=\infty$. Given any surface diffeomorphism $f\colon S_1\to S_2$, the composition map $\widetilde{f}:=\phi_2\circ f\circ\phi_1^{-1}\colon\mathbb{S}^2 \to \mathbb{S}^2$ is a diffeomorphism from $\mathbb{S}^2$ to itself fixing $0$, $1$ and $\infty$. By Theorem \ref{thm:Beltrami}, $\widetilde{f}$ can be uniquely represented by a Beltrami coefficient $\widetilde{\mu}$ defined on $\mathbb{S}^2$. Hence, $f$ can be uniquely represented by a Beltrami coefficient $\mu := \widetilde{\mu}\circ \phi_1 ^{-1}$ defined on $S_1$. In other words, we have the following:

\bigskip

\begin{thm}\label{Cor1}
Let $S_1$ and $S_2$ be two genus-0 closed surfaces. Suppose $f\colon S_1\to S_2$ is a surface diffeomorphism. Given 3-point correspondence $\{p_1, p_2, p_3 \in S_1\} \leftrightarrow \{f(p_1), f(p_2), f(p_3) \in S_2\}$, $f$ can be represented by a unique Beltrami coefficient $\mu : S_1 \to \mathbb{C}$.
\end{thm}

Similarly, let $M_1$ and $M_2$ be two genus-0 open surfaces. Given 2-points correspondence $\{p_1 \in M_1, p_2\in \partial M_1\} \leftrightarrow \{q_1 \in M_2, q_2\in \partial M_2\}$ between them, we can again uniquely parameterize $M_1$ and $M_2$ to map the corresponding points to $0$ and $1$. Denote them by $\phi_1\colon M_1 \to \mathbb{D}$ and $\phi_2\colon M_2 \to \mathbb{D}$. The composition map $\widetilde{f}:=\phi_2\circ f\circ\phi_1^{-1}\colon \mathbb{D} \to \mathbb{D}$ is a diffeomorphism of $\mathbb{D}$ fixing $0$ and $1$. Again, $\widetilde{f}$ can be uniquely represented by a Beltrami coefficient $\widetilde{\mu}$ defined on $\mathbb{D}$. Hence, $f$ can be uniquely represented by a Beltrami coefficient $\mu := \widetilde{\mu}\circ \phi_1 ^{-1}$ defined on $M_1$. So, we have the following theorem:

\bigskip

\begin{thm}\label{Cor2}
Let $M_1$ and $M_2$ be two genus 0 open surfaces with disk topology. Suppose $f\colon M_1\to M_2$ is a surface diffeomorphism. Given 2-point correspondence $\{p_1 \in M_1, p_2\in \partial M_1\} \leftrightarrow \{f(p_1) \in M_2, f(p_2)\in \partial M_2\}$, $f$ can be represented by a unique Beltrami coefficient $\mu\colon M_1 \to \mathbb{C}$.
\end{thm}

Theorem \ref{Cor1} and \ref{Cor2} allow us to represent a homeomorphism of genus-0 closed surfaces or open surfaces with disk topology using a Beltrami coefficient. The Beltrami coefficient unqiuely determine the surface map with least constraint. Using the Beltrami coefficient, we can easily compress the surface maps. These two theorems play important roles for the main algorithms proposed in this paper.

\begin{figure*}[t]
\centering
\includegraphics[height=1.2in]{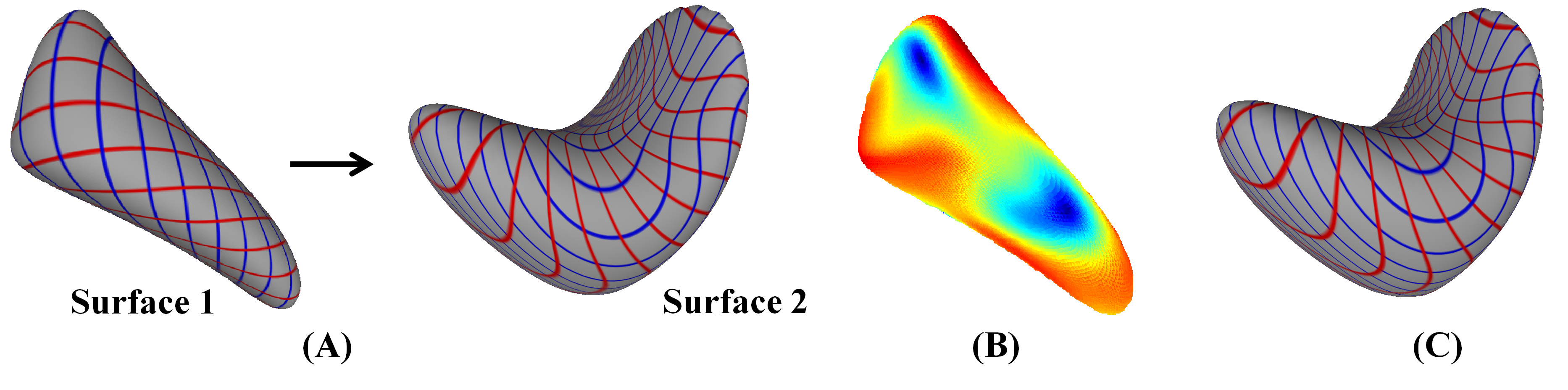}
\caption{Reconstruction of surface homeomorphism from its Beltrami representation. (A) shows the original surface homeomorphism between two genus-0 closed surfaces. (B) shows the reconstructed surface map from its Beltrami representation. The reconstructed map closely resemble to the original one. (C) shows the colormap of the norm of the Beltrami representation. \label{fig:brainreconstructed}}
\end{figure*}

\section{Main Algorithms}

In this section, we describe in detail the main algorithms proposed in this paper to represent and compress surface homeomorphisms using the Beltrami representation. We also describe how the quasi-conformal homeomorphism can be reconstructed from its associated Beltrami representation using the proposed Linear Beltrami Solver.

\subsection{Beltrami representation for bijective maps between meshes}
Surface registration and parameterization are commonly represented by 3D coordinate functions. This representation requires lots of storage space and is difficult to manipulate. For example, the 3D coordinate functions have to satisfy certain constraint on the Jacobian $J$ (namely, $J>0$) in order to preserve the 1-1 correspondence of the surface maps. Enforcing this constraint adds extra difficulty in manipulating or compressing surface maps. It is therefore important to have a simpler representation with as few constraints as possible.

Theorem \ref{Cor1} and \ref{Cor2} tell us that we can represent diffeomorphisms of smooth Riemann surfaces using Beltrami coefficients. In practice, surfaces are represented discretely by triangular meshes. Surface maps are usually approximated by piecewise linear homeomorphisms between meshes. Our goal is to represent piecewise linear maps between meshes by a simple and easy-to-manipulate representation.

Suppose $K_1$ and $K_2$ are two surface meshes with the same topology (either genus-0 closed surface mesh or simply-connected open surface mesh). Define the set of vertices on $K_1$ and $K_2$ by $V^1 = \{v_i^1\}_{i=1}^n$ and $V^2 = \{v_i^2\}_{i=1}^n$ respectively. Similarly, define the set of triangular faces on $K_1$ and $K_2$ by $F^1 = \{T_j^1\}_{j=1}^m$ and $F^2 = \{T_j^2\}_{j=1}^m$ respectively. Now, consider a piecewise linear homeomorphism $f:K_1 \to K_2$ between $K_1$ and $K_2$, which is orientation preserving. In other words, $f|_{T_j^1}$ is piecewise linear and orientation preserving. We aim to represent $f$ by a complex-valued function defined on each triangular face of $K_1$.

\begin{figure*}[t]
\centering
\includegraphics[height=2.25in]{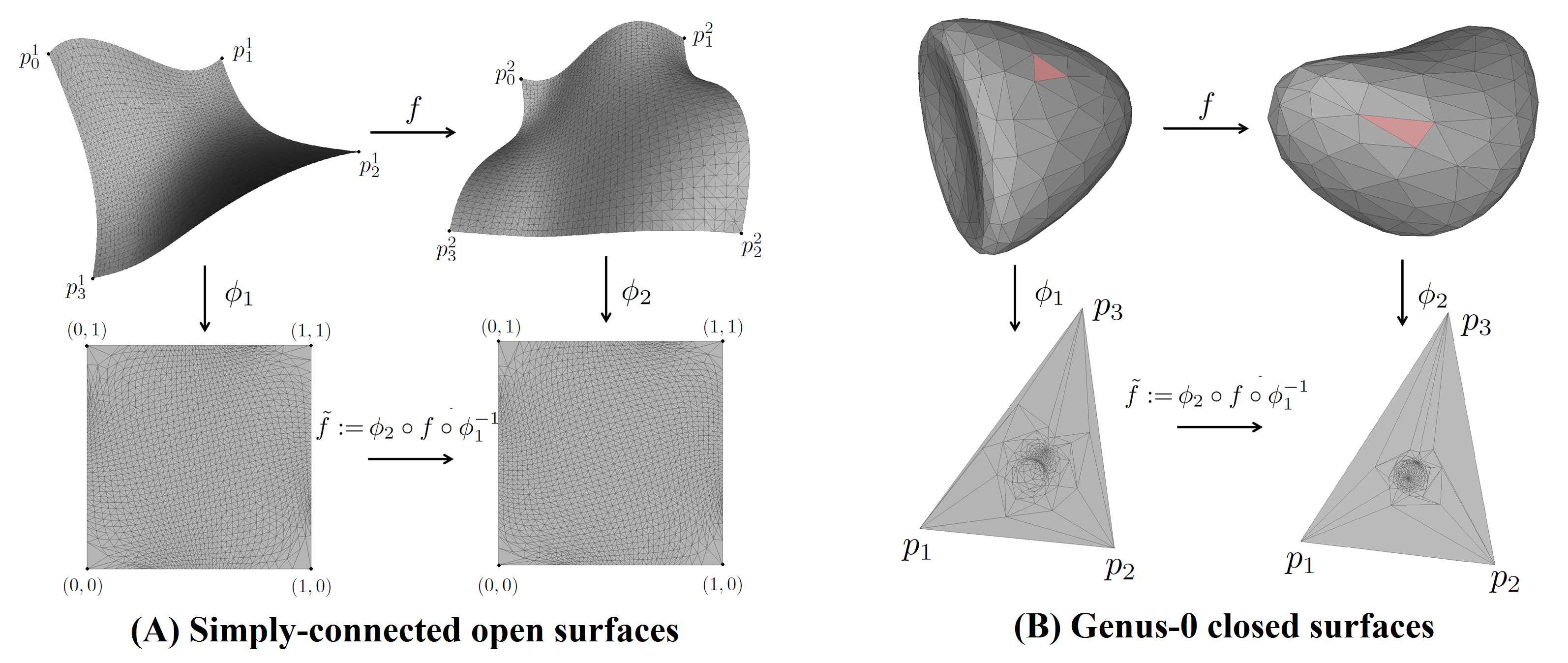}
\caption{Illustration of how Beltrami representations for homeomorphisms between meshes can be computed. (A) shows the case of a homeomorphism between simply-connected open meshes. The two meshes are mapped to a unit square by harmonic parameterizations. (B) shows the case of a homeomorphism between genus-0 closed surface meshes. The two meshes are parameterized onto a triangle in $\mathbb{R}^2$, after cutting away a triangular face on each mesh. \label{fig:openclosed_surface_renew}}
\end{figure*}

We first parameterize $K_1$ and $K_2$. Suppose $K_1$ is a simply-connected open surface mesh. $K_1$ and $K_2$ can be parameterized onto a unit square $R = [0,1]\times [0,1]$ by harmonic maps: $\phi_1: K_1\to R$ and $\phi_2: K_2\to R$ respectively (See Figure \ref{fig:openclosed_surface_renew}(A)). The harmonic parameterizations $\phi_i$ ($i=1,2$) can be computed by solving a sparse linear system \cite{book}:
\begin{equation}\label{eqt:harmonic1}
\begin{split}
& \sum_{[u,v]\in K_i} k_{u,v} (\phi_i(u) - \phi_i (v)) = 0\ \ \ \mathrm{with}\\
\phi_i(p_0) = (0,0);\ & \phi_i(p_1) = (0,1);\ \phi_i(p_2) = (1,0);\ \phi_i(p_3) = (1,1);
\end{split}
\end{equation}

\noindent where $p_0, p_1, p_2$ and $p_3$ are four fixed vertices on the boundary. Other boundary vertices are constrained to be uniformly distributed on the boundary of the unit square $R$. The weight $k_{uv}$ is given by the cotangent formula: $k_{uv} = \cot \alpha^{uv} + \cot \beta^{uv}$ where $\alpha^{uv}$ and $\beta^{uv}$ are the two adjacent angles of the edge $[u,v]$.

In case $K_1$ is a genus-0 closed surface mesh, the surface mesh can be parameterized onto a big triangle on $\mathbb{R}^2$ by cutting away one triangular face (See Figure \ref{fig:openclosed_surface_renew}(B)). The harmonic parameterizations $\phi_i$ ($i=1,2$) from $K_i$ to a triangle on $\mathbb{R}^2$ can be computed as follows:
\begin{equation}\label{eqt:harmonic2}
\begin{split}
\sum_{[u,v]\in K_i} k_{u,v} (\phi_i(u) - \phi_i (v)) = 0\ \ \ \mathrm{with}\\
\phi_i(v_{1}) = p_1,\ \phi_i(v_{2}) = p_2,\ \phi_i(v_{3}) = p_3
\end{split}
\end{equation}

\noindent where $T^i = [v_1^i, v_2^i, v_3^i]$ corresponds to the cut away triangular face on $K_i$ and $\{p_1, p_2, p_3\}$ are three fixed point on $\mathbb{R}^2$.

The bijective map $f:K_1\to K_2$ can now be represented by the Beltrami coefficient $\mu_{\tilde{f}}$ of the composition map $\tilde{f}: \phi_2\circ f\circ\phi_1 ^{-1}:D \to D$, where $D$ is a triangular mesh in $\mathbb{R}^2$. $D$ is either the unit square or a triangle in the complex plane (See Figure \ref{fig:openclosed_surface_renew}).

To compute $\mu_{\tilde{f}}$, we simply need to approximate the partial derivatives at each face $T$. We denote them by $D_x \tilde{f}(T)$ and  $D_y \tilde{f}(T)$ respectively. Note that $\tilde{f}$ is piecewise linear. The restriction of $\tilde{f}$ on each triangular face $T$ can be written as:
\begin{equation}
\tilde{f}|_T (x,y) = \left( \begin{array}{c}
a_T x + b_T y + r_T \\
c_T x + d_T y + q_T \end{array} \right)
\end{equation}

Hence, $D_x \tilde{f}(T) = a_T + i c_T$ and $D_y \tilde{f}(T) = b_T + i d_T$. Now, the gradient $\nabla _T \tilde{f} := (D_x \tilde{f}(T), D_y \tilde{f}(T))^t$ on each face $T$ can be computed by solving:
\begin{equation}\label{eqt:gradient}
\left( \begin{array}{c}
\vec{v}_1 - \vec{v}_0\\
\vec{v}_2 - \vec{v}_0\end{array} \right)\nabla_T \tilde{f}_i = \left( \begin{array}{c}
\tilde{f}_i(\vec{v}_1) - \tilde{f}_i(\vec{v}_0)\\
\tilde{f}_i(\vec{v}_2) - \tilde{f}_i(\vec{v}_0)\end{array} \right),
\end{equation}

\noindent where $[\vec{v_0},\vec{v_1}]$ and $[\vec{v_0},\vec{v_2}]$ are two edges on $T$. By solving equation \ref{eqt:gradient}, $a_T$, $b_T$, $c_T$ and $d_T$ can be obtained. The Beltrami coefficient $\mu_{\tilde{f}}(T)$ of the triangular face $T$ can then be computed from the Beltrami equation \ref{beltramieqt} by:
\begin{equation}\label{eqt:BC}
\mu_{\tilde{f}}(T) = \frac{(a_T - d_T)+\sqrt{-1}(c_T + b_T)}{(a_T + d_T)+\sqrt{-1}(c_T - b_T)}
\end{equation}

The Beltrami coefficient $\mu_{\tilde{f}}$ defined on each triangular face of $D$ uniquely determine the surface map $\tilde{f}$. Hence, the complex-valued function $\mu_f := \mu_{\tilde{f}}\circ \phi_1$ defined on each face of $K_1$ uniquely determine the surface map $f:K_1\to K_2$. We call $\mu_f$ the {\it Beltrami representation} of $f$. Given a Beltrami representation, the corresponding surface homeomorphism can be exactly computed, which will be described in the next subsection (See Figure \ref{fig:xyexample} and \ref{fig:brainreconstructed}).

In summary, the Beltrami representation of a homeomorphism between surface meshes can be computed as follows.

\bigskip

\noindent $\mathbf{Algorithm\ 4.1:}$ {\it(Computation of Beltrami representation)}\\
\noindent $\mathbf{Input:}$ {\it Surface meshes: $K_1$ and $K_2$; piecewise linear homeomorphism: $f:K_1\to K_2$; a fixed triangular face $T$ if $K_1$ is genus-0 or 4 fixed points on $\partial K_1$ if $K_1$ is a topological disk.}\\
\noindent $\mathbf{Output:}$ {\it Beltrami representation: $\mu_f$}\\
\vspace{-3mm}
\begin{enumerate}
\item {\it Get harmonic parameterizations $\phi_1: K_1\to D$ and $\phi_2:K_2\to D$ by solving the linear system \ref{eqt:harmonic1} or \ref{eqt:harmonic2}}
\item {\it Compute the gradient of $\tilde{f}:= \phi_2\circ f\circ\phi_1 ^{-1}:D \to D$ for each triangular face of $D$ by solving the linear system \ref{eqt:gradient}}
\item {\it Compute the Beltrami coefficient of $\tilde{f}$ for each triangular face of $D$ by equation \ref{eqt:BC}}
\item {\it Compute the Beltrami representation $\mu_f$ of $f$ by computing $\mu_f := \mu_{\tilde{f}}\circ \phi_1$.}
\end{enumerate}

\bigskip

\subsection{Linear Beltrami solver}
Given a Beltrami representation, it is important to have an algorithm to reconstruct the associated quasi-conformal homeomorphism.

Suppose $K_1$ and $K_2$ are two surface meshes with the same topology (either topological disks or genus-0 closed surface meshes). Given a Beltrami representation $\mu$, which is a complex-valued function defined on each triangular face of $K_1$, our goal is to reconstruct the associated homeomorphism $f:K_1\to K_2$ between $K_1$ and $K_2$.

We first compute the harmonic parameterizations $\phi_1:K_1\to D$ and $\phi_2:K_2\to D$ using equation \ref{eqt:harmonic1} or \ref{eqt:harmonic2}. We shall look for a piecewise linear homeomorphism $\tilde{f}:D\to D$ with Beltrami coefficient $\mu_{\tilde{f}} : = \phi_1\circ \mu$. The desired homeomorphism $f$ can then be reconstructed by taking the composition: $f = \phi_2 \circ \tilde{f} \circ \phi_1^{-1}$. Again, the restriction of $\tilde{f}$ on each triangular face $T$ is linear and thus it can be written as:
\begin{equation}
\tilde{f}|_T (x,y) = \left( \begin{array}{c}
a_T x + b_T y + r_T \\
c_T x + d_T y + q_T \end{array} \right)
\end{equation}

Let $\mu(T) = \rho_T + \sqrt{-1}\ \tau_T$. Equation \ref{eqt:BC} gives

\begin{equation}
\rho_T + \sqrt{-1}\ \tau_T = \frac{(a_T - d_T)+\sqrt{-1}(c_T + b_T)}{(a_T + d_T)+\sqrt{-1}(c_T - b_T)}
\end{equation}

This is equivalent to
\begin{equation}\label{eqt:BCsplit1}
\begin{split}
-d_T & = \alpha_1^T a_T + \alpha_2^T b_T\\
 c_T & = \alpha_2^T a_T + \alpha_3^T b_T
\end{split}
\end{equation}

\noindent where:
$\alpha_1^T = \frac{(\rho_T -1)^2 + \tau_T^2}{1-\rho_T^2 - \tau_T^2} $; $\alpha_2^T = -\frac{2\tau_T}{1-\rho_T^2 - \tau_T^2} $; $\alpha_3^T = \frac{1+2\rho_T+\rho_T^2 +\tau_T^2}{1-\rho_T^2 - \tau_T^2} $.

Similarly, one can deduce that
\begin{equation}\label{eqt:BCsplit2}
\begin{split}
-b_T & = \alpha_1^T c_T + \alpha_2^T d_T\\
 a_T & = \alpha_2^T c_T + \alpha_3^T d_T
\end{split}
\end{equation}

Let $T = [v_i,v_j, v_k]$ and $w_I = \tilde{f}(v_I)$ where $I=i,j$ or $k$. Suppose $v_I = g_I + \sqrt{-1}\ h_I$ and $w_I = s_I + \sqrt{-1}\ t_I$ ($I=i,j,k$). Using equation \ref{eqt:gradient}, $a_T, b_T, c_T$ and $d_T$ can be written as follows:
\begin{equation}
\begin{split}
a_T = A_i^T s_i + A_j^T s_j + A_k^T s_k;\ b_T = B_i^T s_i + B_j^T s_j + B_k^T s_k;\\
c_T = A_i^T t_i + A_j^T t_j + A_k^T t_k;\ d_T = B_i^T t_i + B_j^T t_j + B_k^T t_k;
\end{split}
\end{equation}
\noindent where:
\begin{equation}
\begin{split}
&A_i^T = (h_j-h_k )/2Area(T),\ A_j^T = (h_k-h_i )/2Area(T),\ A_k^T = (h_i-h_j )/2Area(T);\\
&B_i^T = (g_k-g_j )/2Area(T),\ B_j^T = (g_i-g_k )/2Area(T),\ B_k^T = (g_j-g_i )/2Area(T);
\end{split}
\end{equation}

For each vertex $v_i$, let $N_i$ be the collection of neighborhood faces attached to $v_i$. By careful checking, one can observe that
\begin{equation}
\sum_{T\in N_i} A_i^T b_T Area(T)= \sum_{T\in N_i} B_i^T a_T Area(T);\ \sum_{T\in N_i} A_i^T d_T Area(T)= \sum_{T\in N_i} B_i^T c_T Area(T).
\end{equation}

Thus, following from equation \ref{eqt:BCsplit1} and \ref{eqt:BCsplit2}, we have
\begin{equation}\label{eqt:linearB1}
\sum_{T\in N_i} \left(A_i^T [\alpha_1^T a_T + \alpha_2^T b_T] + B_i^T[\alpha_2^T a_T + \alpha_3^T b_T]\right)Area(T) = 0
\end{equation}

\begin{equation}\label{eqt:linearB2}
\sum_{T\in N_i} \left(A_i^T [\alpha_1^T c_T + \alpha_2^T d_T] + B_i^T[\alpha_2^T c_T + \alpha_3^T d_T]\right)Area(T) = 0
\end{equation}

\noindent for all vertices $v_i \in D$. Note that $a_T$ and $b_T$ can be written as a linear combination of the x-coordinates of the desired quasi-conformal map $\tilde{f}$. Hence, equation \ref{eqt:linearB1} gives us the linear systems to solve for the x-coordinate function of $\tilde{f}$. Similarly, $c_T$ and $d_T$ can also be written as a linear combination of the y-coordinates of the desired quasi-conformal map $\tilde{f}$. Therefore, equation \ref{eqt:linearB2} gives us the linear systems to solve for the y-coordinate function of $\tilde{f}$.

Besides, $\tilde{f}$ has to satisfy certain constraints on the boundary. When $K_1$ is a genus-0 closed surface mesh, the parameter domain $D$ is a triangle with boundary vertices $p_0$, $p_1$ and $p_2$. In this case, the desired quasi-conformal map $\tilde{f}$ should satisfy
\begin{equation}\label{eqt:bounday1}
\tilde{f}(p_0) = p_0; \tilde{f}(p_1) = p_1\ \mathrm{and\ }\tilde{f}(p_2) = p_2
\end{equation}

When $K_1$ is a topological disk, the parameter domain $D$ is unit square. In this case, the desired quasi-conformal map should satisfy
\begin{equation}\label{eqt:bounday2}
\begin{split}
\tilde{f}(0) = 0; \tilde{f}(1) = 1\ \tilde{f}(i) = i\ \tilde{f}(1+i) = 1+i;\\
\mathbf{Re}(\tilde{f}) = 0 \mathrm{\ on\ arc\ }[0, i];\ \mathbf{Re}(\tilde{f}) = 1 \mathrm{\ on\ arc\ }[1, 1+i];\\
\mathbf{Imag}(\tilde{f}) = 0 \mathrm{\ on\ arc\ }[0, 1];\ \mathbf{Imag}(\tilde{f}) = 1 \mathrm{\ on\ arc\ }[i, 1+i]
\end{split}
\end{equation}

Equations \ref{eqt:linearB1} and \ref{eqt:linearB2} together with the above boundary conditions give a non-singular linear system to solve for $\tilde{f}$. The linear system is symmetric positive definite. Hence, it can be solved effectively by the conjugate gradient method. Once $\tilde{f}$ is computed, the desired quasi-conformal map between $K_1$ and $K_2$ can be computed by taking the composition: $f = \phi_2^{-1} \circ \tilde{f} \circ \phi_1$.

We summarize the reconstruction scheme of the quasi-conformal map from its Beltrami representation as follows.

\medskip

\noindent $\mathbf{Algorithm\ 4.2:}$ {\it(Linear Beltrami Solver)}\\
\noindent $\mathbf{Input:}$ {\it Surface meshes: $K_1$ and $K_2$; Beltrami representation $\mu$}\\
\noindent $\mathbf{Output:}$ {\it Quasi-conformal map $f:K_1\to K_2$ associated to $\mu$}\\
\vspace{-3mm}
\begin{enumerate}
\item {\it Get harmonic parameterizations $\phi_1: K_1\to D$ and $\phi_2:K_2\to D$ by solving the linear system \ref{eqt:harmonic1} or \ref{eqt:harmonic2}}
\item {\it Compute the Beltrami coefficient  $\mu_{\tilde{f}} : = \mu \circ \phi_1^{-1}$}
\item {\it Compute the quasi-conformal map $\tilde{f}:D\to D$ associated to $\mu_{\tilde{f}}$ by solving the linear system \ref{eqt:linearB1} and \ref{eqt:linearB2} with boundary conditions \ref{eqt:bounday1} or \ref{eqt:bounday2}}
\item {\it Compute the quasi-conformal map $f:K_1\to K_2$ by $f = \phi_2^{-1} \circ \tilde{f} \circ \phi_1$}
\end{enumerate}

\medskip

In some situation, the surface mesh may be parameterized onto an arbitrary domain $\Omega$ in $\mathbb{R}^2$ (such as in texture mapping applications). In this case, the Beltrami representation together with the boundary map uniquely determines the parametrization. By solving Equation \ref{eqt:linearB1} and \ref{eqt:linearB2} together with the boundary condition given by the boundary map, the parameterization can be exactly reconstructed.

Experimental results show that the proposed Linear Beltrami Solver can effectively compute the surface homeomorphism associated to a given Beltrami representation. Compared to the Beltrami Holomorphic Flow (BHF) method introduced in \cite{LuiCompression,LuiBHF}, our method can compute the associated surface homeomorphism more accurately and efficiently. Table \ref{compareLBSBHF} shows the comparison of the computational time and error under the LBS method and the BHF method. Experimental results show that LBS method can compute the associated surface map much faster than the BHF method. Also, the accuracy is much better when using LBS. As shown in Table \ref{compareLBSBHF}, the numerical error under LBS is much less than that under BHF. The computational efficiency of LBS allows us to apply our proposed algorithm in more practical applications that require real-time processing, such as video compression.

\begin{table}
\caption{\label{compareLBSBHF} Comparison of the computational time and error under LBS and BHF.}
\vspace{-5mm}
\begin{center}
\begin{tabular}{c||c|c|c|c|c}
&$||\mu||_{\infty}$& L1-error (LBS) & Time (LBS) & L1-error (BHF) & Time (BHF) \\
\hline
Map 1 & 0.6258 & $1.365\times 10^{-14}$&  0.032 s & 0.0064 & 98.5 s \\
Map 2 & 0.8556 & $9.326\times 10^{-15}$&  0.028 s & 0.0159 & 96.4 s\\
Map 3 & 0.9995 & $2.408\times 10^{-13}$ & 0.033 s & 0.0313 & 102.2 s\\
\end{tabular}
\end{center}
\end{table}

\bigskip

\subsection{Compression of surface homeomorphisms}
The Beltrami representation can be further compressed using the Fourier approximation to reduce the storage requirement. An important consideration is to preserve the bijectivity of the reconstructed surface map after the compression.

\begin{figure*}[t]
\centering
\includegraphics[height=1.60in]{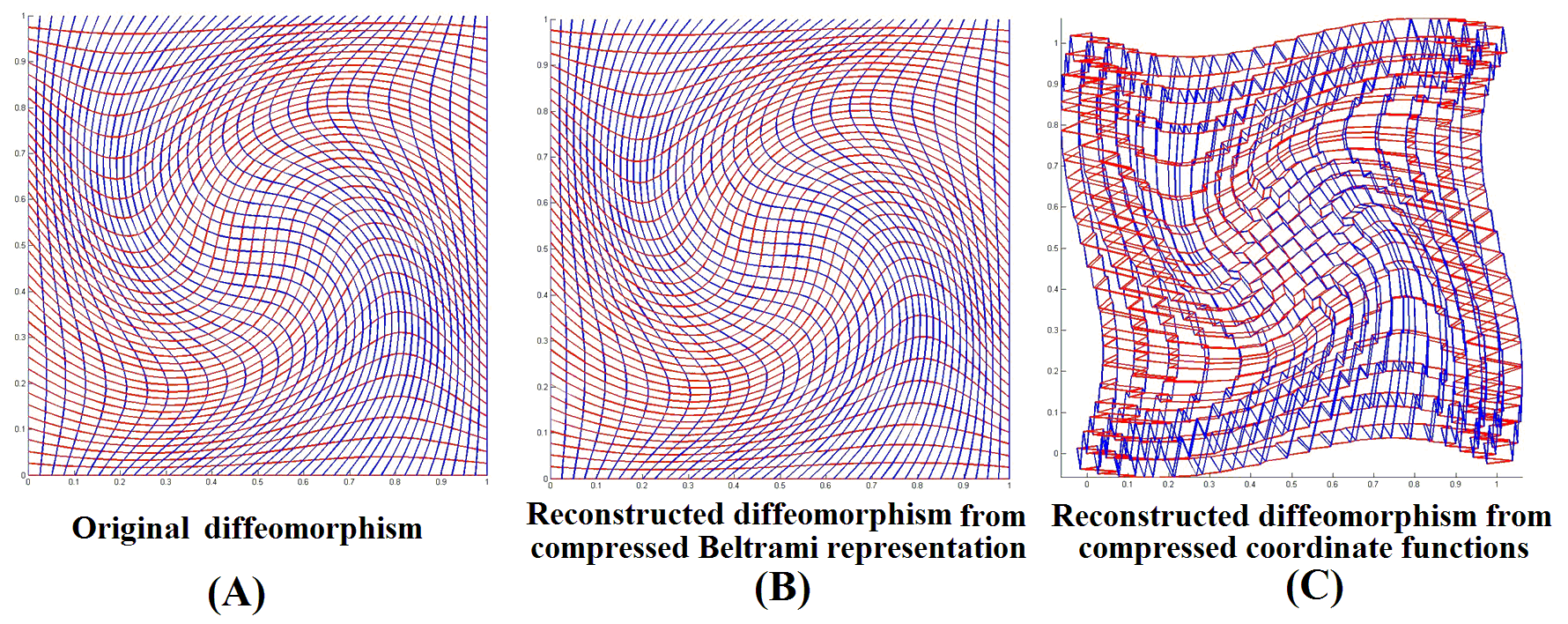}
\caption{Compression of bijective map. (A) shows the original bijective map of the unit square. (B) shows the reconstructed map from the compressed Beltrami representation. The reconstructed map closely resembles to the original map in (A). (C) shows the reconstructed map from the compressed coordinate functions. The bijectivity of the map is completely disrupted. \label{fig:Grid_xy_fail}}
\end{figure*}

\begin{figure*}[ht]
\centering
\includegraphics[height=3.41in]{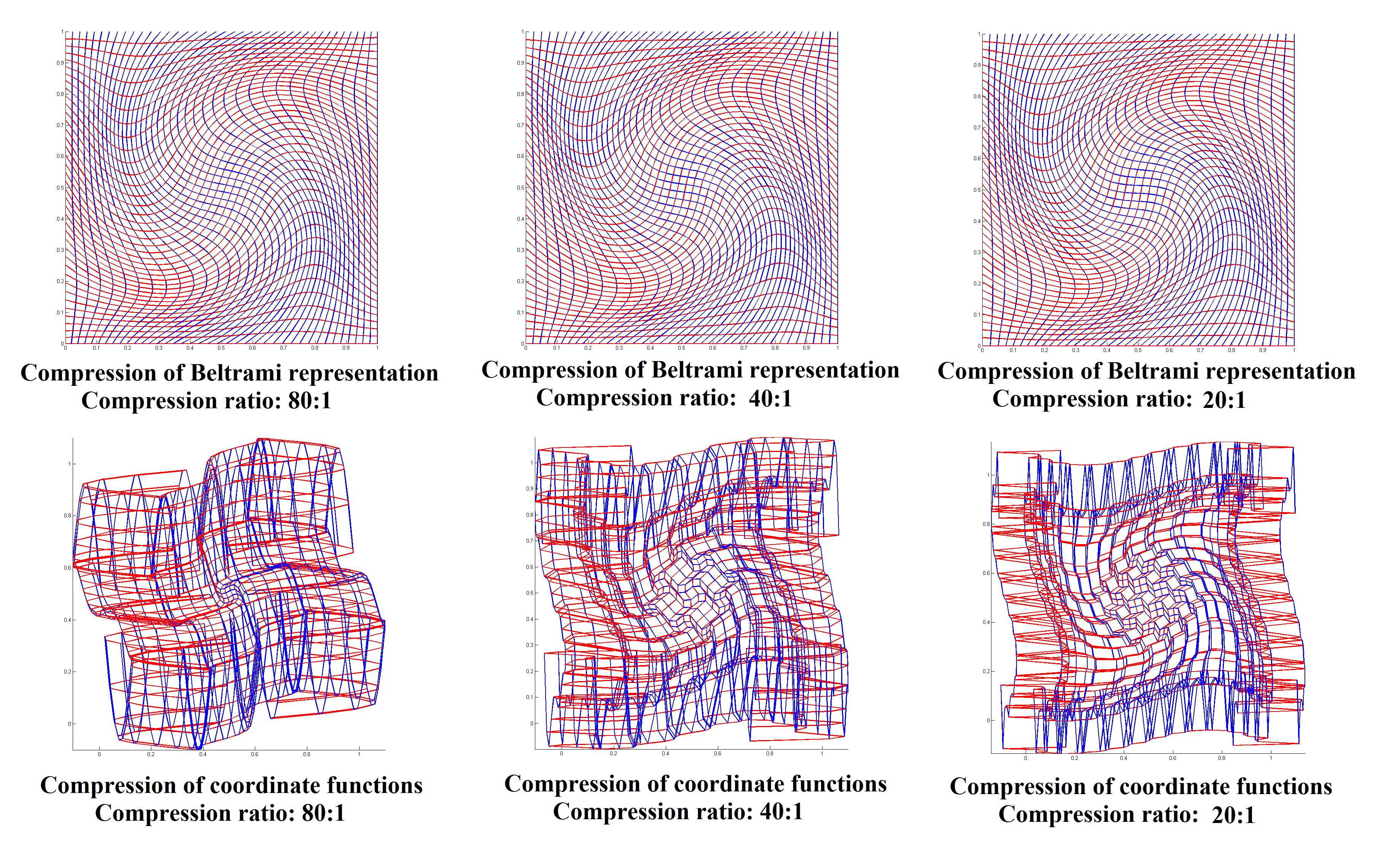}
\caption{Compression of bijective map with different compression ratios. The top row shows the reconstructed map from the compressed Beltrami representation with different compression ratios. The bottom row shows the reconstructed map from the compressed coordinate functions with the same compression ratios as in the top row. \label{fig:BCcoordinatecompare}}
\end{figure*}

\begin{figure*}[t]
\centering
\includegraphics[height=1.5in]{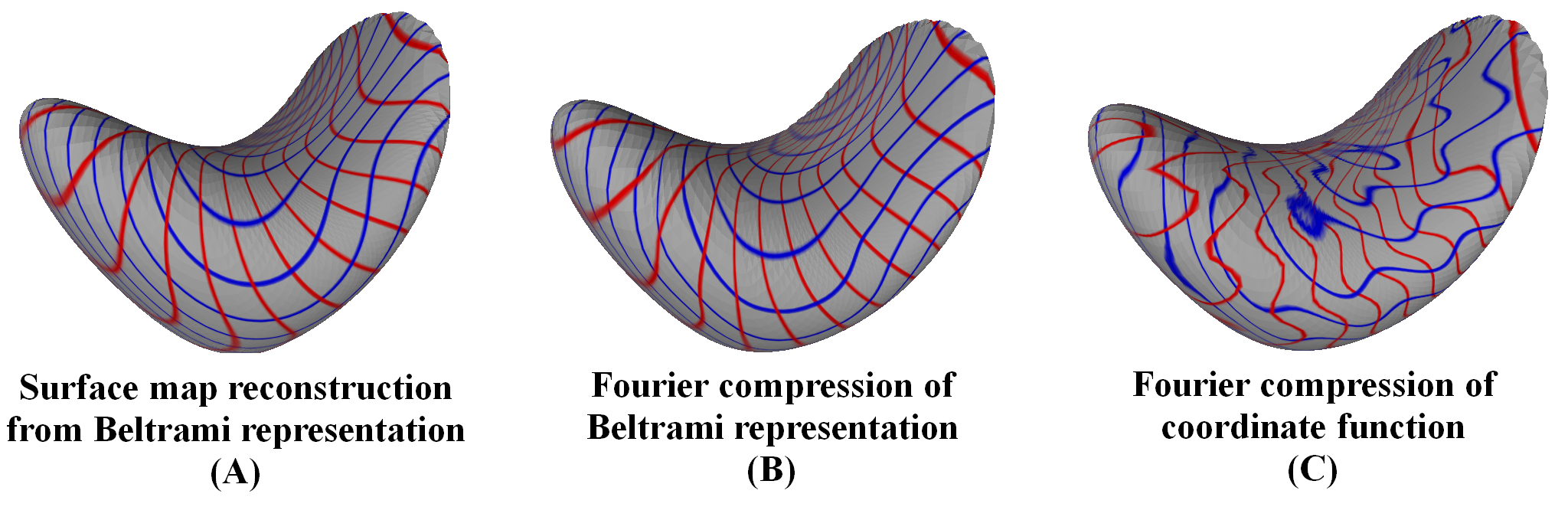}
\caption{Compression of bijective surface homeomorphism between two genus-0 closed surfaces. (A) shows the original surface homeomorphism between two closed surfaces. (B) shows the reconstructed surface map from the compressed Beltrami representation. (C) shows the the reconstructed surface map from the compressed coordinate functions. \label{fig:brainfourier}}
\end{figure*}

Under the representation by coordinate functions, the surface map cannot be easily compressed using the Fourier approximation without distorting the bijectivity. In order to preserve the bijectivity, the Jacobian of the coordinate functions has to be greater than 0. This constraint is equivalent to an inequality in the partial derivatives of the coordinate functions. Enforcing this constraint is difficult during compression and the bijective property is easily lost (see Figure \ref{fig:Grid_xy_fail}, \ref{fig:BCcoordinatecompare} and \ref{fig:brainfourier}). The Beltrami representation, however, is advantageous because it does not have any requirement for injectivity and surjectivity, making the Jacobian constraint unnecessary. The only requirement for the Beltrami representation $\mu$ is that it has to be a complex-valued function defined on the triangular faces with supreme norm less than 1. We can therefore compress $\mu$ using Fourier approximations without losing the bijectivity.

Let $K_1$ be a simply-connected open surface mesh. Suppose $\mu$ is the Beltrami representation of the surface map $f:K_1 \to K_2$. We first parameterize $K_1$ to the unit square $D$ by the parameterization $\phi_1: K_1\to D$. We can then obtain the Beltrami coefficient $\mu_{\tilde{f}}:=\mu \circ \phi_1^{-1}$, which is defined on triangular faces of $D$. Let $D_{reg}$ be the $N\times N$ regular finite difference grid of the unit square. Here, $N = \sqrt{N_f}$, where $N_f$ is the number of triangular faces of $K_1$. We linearly interpolate $\mu_{\tilde{f}}$ defined on $D$ onto $D_{reg}$ to obtain a Beltrami coefficient $\mu^{reg}_{\tilde{f}}$ defined on $D_{reg}$.

$\mu^{reg}_{\tilde{f}}$ can be expressed by Fourier expansion as follow:
\begin{equation}\label{eqt:Fourier}
\mu (x,y) = \sum_{j=0}^{N-1} \sum_{k=0}^{N-1} c_{j,k} e^{2\sqrt{-1}\pi jx/N}e^{2\sqrt{-1}\pi ky/N},
\end{equation}
\noindent where
\begin{equation}\nonumber
c_{j,k} = \frac{1}{N^2} \sum_{x=0}^{N-1} \sum_{y=0}^{N-1} \mu (x,y)e^{-2\sqrt{-1}\pi jx/N}e^{-2\sqrt{-1}\pi ky/N}.
\end{equation}

We can use fast Fourier transform to compute the coefficients $c_{j,k}$ efficiently. One can then take fewer Fourier coefficients to approximate the Beltrami coefficient, which can significantly reduce the storage requirement. With the Fourier coefficients of the truncated Fourier series, the compressed Beltrami coefficient defined on $D$ and hence the compressed Beltrami representation $\mu^{c}$ defined on $K_1$ can be obtained. $\mu^{c}$ accurately approximates $\mu$, and hence the quasi-conformal map associated to $\mu^{c}$ closely resembles to $f$.

In summary, the Fourier compression scheme can be described as follows:

\bigskip

\noindent $\mathbf{Algorithm\ 4.3:}$ {\it(Compression of Beltrami representation)}\\
\noindent $\mathbf{Input:}$ {\it Beltrami representation $\mu$ and compression percentage $\epsilon$}\\
\noindent $\mathbf{Output:}$ {\it $\epsilon \%$ Fourier coefficients $c_{j,k}$'s}\\
\vspace{-3mm}
\begin{enumerate}
\item {\it Get harmonic parameterization $\phi_1: K_1\to D$ by solving the linear system \ref{eqt:harmonic2}}
\item {\it Compute the Beltrami coefficient  $\mu_{\tilde{f}} : = \phi_1\circ \mu$}
\item {\it Compute $\mu^{reg}_{\tilde{f}}$ defined on the regular grid $D_{reg}$ by linear interpolation}
\item {\it Compute the Fourier coefficients of $\mu^{reg}_{\tilde{f}}$ using equation \ref{eqt:Fourier}}
\item {\it Store $\epsilon \%$ Fourier coefficients $c_{j,k}$'s}
\end{enumerate}

\bigskip

Experimental results show that the proposed compression algorithm is stable and effective in reducing the storage requirement of bijective surface maps. Figure \ref{fig:Grid_xy_fail}(A) shows a homeomorphism from a regular grid. (B) shows the reconstructed homeomorphism from the compressed Beltrami representation. The reconstructed map closely resembles to the original one. (C) shows the reconstructed map from the compressed coordinate functions. Note that the bijectivity is completely disrupted. In comparison, Beltrami compression gives accurate results with just a small number of coefficients. Figure \ref{fig:BCcoordinatecompare} shows the result of Beltrami compression with different compression ratios. The reconstructed maps again closely resemble to the original data. In the bottom row of Figure \ref{fig:BCcoordinatecompare}, we show the reconstructed map from the compressed coordinate functions with the same compression ratios. Again, the bijectivity of the reconstructed maps is completely disrupted.

The proposed compression scheme can also be applied to compressing surface homeomorphisms. Figure \ref{fig:brainfourier}(A) shows the original surface homeomorphism between two genus-0 closed surfaces. (B) shows the reconstructed surface map from the compressed Beltrami representation, which closely resemble to the original map in (A). (C) shows the the reconstructed surface map from the compressed coordinate functions. It is observed again that the bijectivity of the surface map cannot be preserved after the compression.

\bigskip

\section{Applications}

\begin{figure}[t]
\centering
\includegraphics[height=3in]{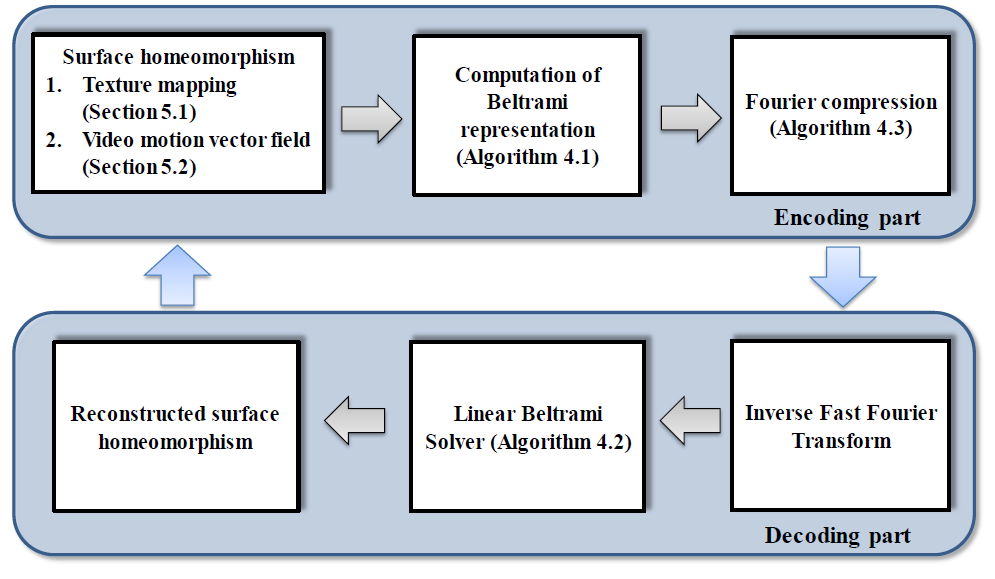}
\caption{Summary of the steps of encoding and decoding of texture mapping/video compression.\label{fig:flow_chart}}
\end{figure}

In this section, we apply our proposed compression algorithm to texture map and video compression. A flow chart summarizing the idea of the two applications is shown in Figure \ref{fig:flow_chart}.

\subsection{Compression of Texture Mapping}

Texture mapping techniques have been extensively studied to provide realistic 3D rendering in movies, animation and video gaming. The basic idea of texture mapping is to map a texture image onto a given surface, so as to increase the realism of the 3D model. As the demand for higher level of realism is rising, texture mapping of higher resolution is also required for producing more detailed models, which in turn increases the required storage capacity and transmission bandwidth for texture coordinates. The need for a compact representation of the texture mapping is therefore crucial. In this section, we propose to develop an effective algorithm for texture map compression using the Beltrami representation.

Mathematically, a texture map is a parameterization of a surface onto a 2D image. This parameterization is commonly represented by its coordinate functions $\rho = (\rho_1, \rho_2)$, which maps each vertex of the mesh to a 2D position, given by 
\begin{equation}
\rho:M \rightarrow [0,1]\times[0,1]
\end{equation}
where $M$ is the surface mesh. Conventionally, the target range is set to be $[0,1]\times[0,1]$, which can be obtained by normalizing the texture image.  In other words, a texture mapping can be considered as a piecewise linear homeomorphism between the surface mesh $M$ and a 2D mesh in $[0,1]\times[0,1]$ . Once every vertices are assigned to a 2D position, the texture image can be mapped onto the 3D surface mesh.

\begin{figure*}[t]
\centering
\includegraphics[height=1.15in]{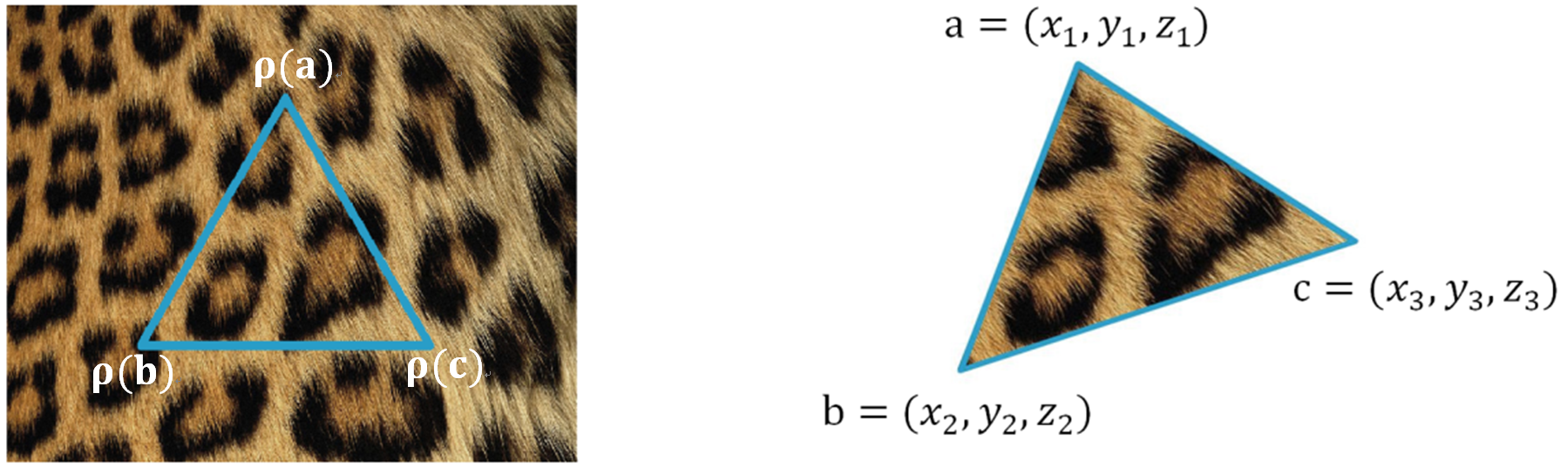}
\caption{Illustration of a typical texture mapping in a triangular mesh. Vertices $a, b$ and $c$ are mapped to $\rho(a), \rho(b)$ and $\rho(c)$ respectively. Texels in the interior of the triangle $\bigtriangleup \rho(a) \rho(b) \rho(c)$ can then be assigned to the interior of $\bigtriangleup abc$. Note that interior angles are changed after the mapping, indicating the quasiconformality of the mapping.\label{fig:texture_explain}}
\end{figure*}

Figure \ref{fig:texture_explain} illustrates the idea of texture mapping. The left shows a 2D texture image. Vertices $a,b$ and $c$ have texture coordinates $\rho(a),\rho(b)$ and $\rho(c)$ respectively. With this per-vertex assignment, texels can be assigned to the interior of the triangle by linear interpolation. In real application, a 3D model is usually represented by fine meshes composed of many small triangular faces. With the texture mapping, texels can be mapped to the interior of each triangular faces, and hence the whole textured surface mesh can be obtained. 

\begin{figure*}[h]
\centering
%\includegraphics[height=1.60in]{figs/buddha_example.png}
%\caption{\label{buddha_example}}
\includegraphics[height=1.65in]{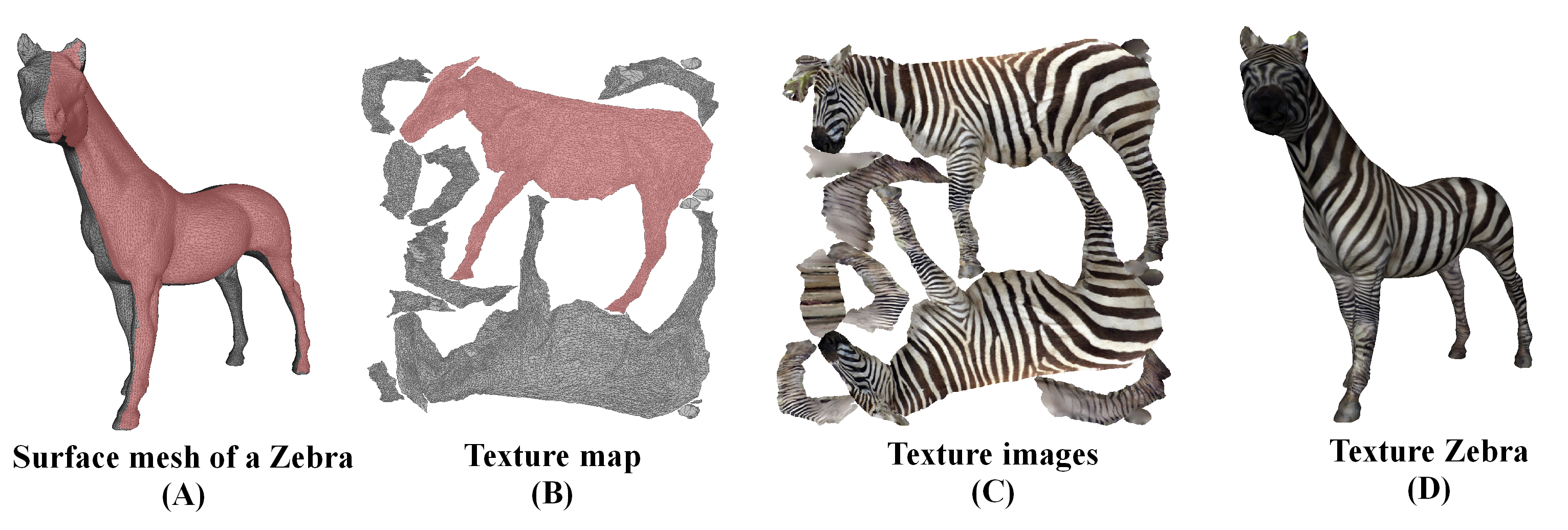}
\caption{\label{Zebra_example} 	Texture images with arbitrary shapes are common in texture mapping techniques. With the texture coordinates, texels of the texture can be assigned to the corresponding position in the mesh. (A)Mesh of Zebra. (B)The texture maps of Zebra. (C)The texture images of all parts of Zebra. (D)Texture mapping result of Zebra. $^1$}
\end{figure*}

Texture images are not necessarily of regular shapes. In order to minimize the distortion of the texture mapping, surface meshes are often partitioned into patches and parameterized onto texture images with irregular shapes. Figure \ref{Zebra_example} shows a typical example of texture mapping.  (A) shows the surface mesh of a zebra, which is partitioned into patches. Each patch is parameterized onto a texture image with irregular shapes, as shown in (B) and (C). In other words, the surface mesh consists of several texture mappings corresponding to different patches. With the texture mappings, texture images can be mapped onto the surface mesh. (D) shows the textured surface mesh of the zebra.

In practice, a textured surface mesh is commonly represented by its mesh geometry, mesh connectivity, texture mapping and texture image. Usually, the storage memories for texture properties contribute a significant portion of the total file size of the textured mesh.  This motivates us to look for an effective compression scheme for texture mapping.

Here, we propose an effective algorithm for texture mapping compression by the Beltrami representation of the texture coordinates.  Consider a surface mesh $M$ with texture mappings $\{\rho_i: M_i \to [0,1]\times[0,1]\}_{i=1}^n$, where $M_i$'s are patches of $M$. Our goal is to compress the texture mappings $\rho_i$'s. We first parameterize $M_i$ onto a unit square $D$ by harmonic parameterizations $\phi_i: M_i\to D$. It can be done by solving Equation \ref{eqt:harmonic1}. The Beltrami coefficient $\tilde{\mu_i}$ of $\tilde{\rho_i} : = \rho\circ\phi_i^{-1}$ can then be computed.   As described in Section 4.1, the Beltrami representation $\mu_i := \tilde{\mu}_i \circ \phi_i$ together with the boundary map $\rho_i |_{\partial M_i}$ uniquely determine $\rho_i$. Using the Fourier compression scheme as described in Algorithm 4.2, the Beltrami representation $\mu_i$ can be effectively compressed.  Hence, the texture mapping $\rho_i$ can now be represented by 1) the Fourier coefficients $c_{j,k}^i$ in the truncated Fourier series, and 2) the texture coordinates of all boundary vertices of $M$. This significantly reduces the storage requirement for the texture mappings. Note that in the case the texture image is of regular shape (e.g. 2D rectangle), storing the texture coordinates of the boundary vertices is not necessary. Instead, texture coordinates of 4 boundary vertices are sufficient.

The decoding algorithm is also straightforward. Given the Fourier coefficients $c_{j,k}^i$  and the texture coordinates  of the boundary vertices. Our goal is to reconstruct the texture mapping $\rho_i$. Using the inverse Fast Fourier transform applied to Fourier coefficients $c_{j,k}^i$  saved, the Beltrami representation $\mu_i :M_i \to \mathbb{C}$ can be restored. With the Beltrami representation, the texture mapping $\rho_i$ can be reconstructed using the Linear Beltrami Solver in Algorithm 4.2. More specifically, $\rho_i$ can be computed by solving the linear systems $\ref{eqt:linearB1}$ and $\ref{eqt:linearB2}$ , subject to the boundary condition given by the texture coordinates of the boundary vertices. Texture coordinates of $M_i$ can then be obtained.

\let\thefootnote\relax\footnotetext{1. The texture mapping example is freely available on http://www.kunzhou.net/tex-models.htm}

\begin{figure*}[t]
\centering
\includegraphics[height=1.55in]{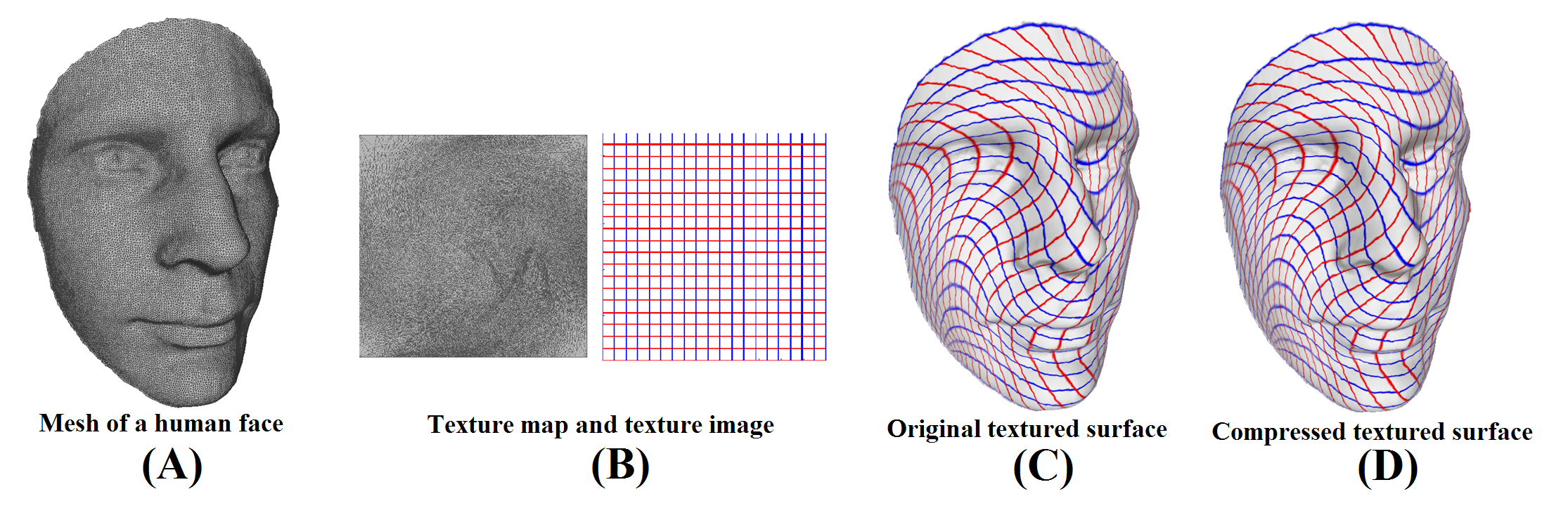}
\caption{Illustration of the proposed compression scheme. No visible difference can be found after the reconstruction.(A)Original mesh of a man face. (B)The corresponding texture map and the texture. (C)The original texture mapping result of the man face. (D)The compressed texture mapping result of the man face. \label{fig:manface}}
\end{figure*}

\begin{figure*}[h!]
\centering
\includegraphics[height=1.65in]{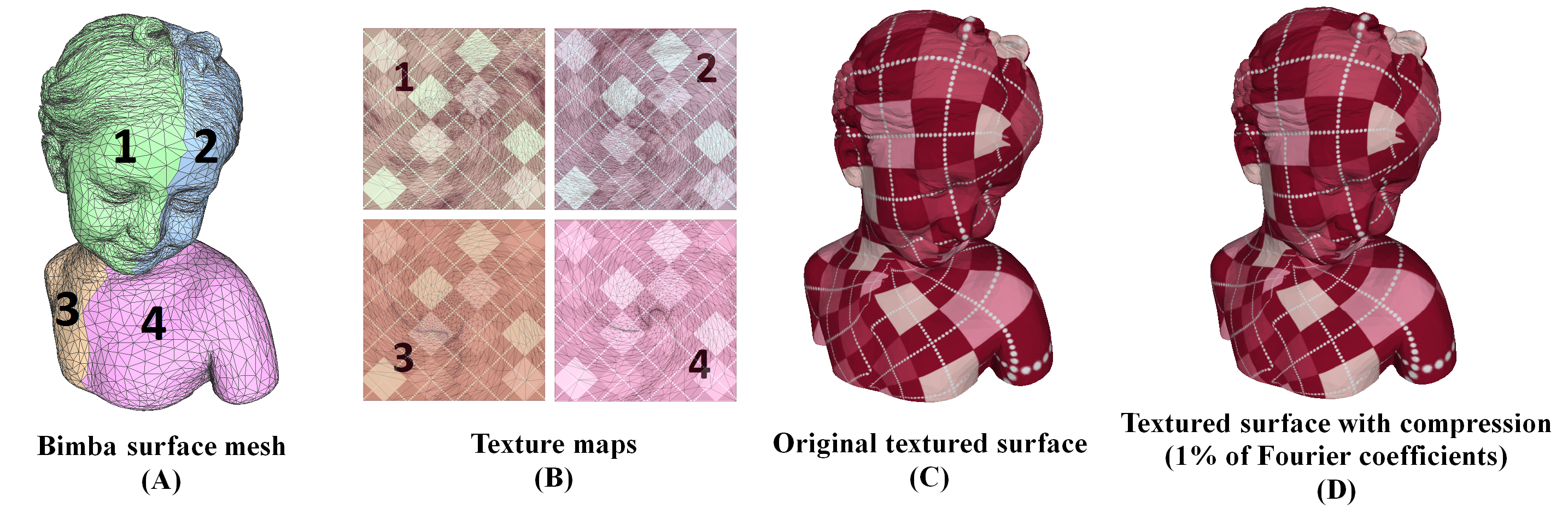}
\caption{Illustration of the proposed compression scheme. The Bimba is first partitioned into four parts. Texture maps are then defined independently. (A)Mesh of the Bimba. (B)The texture map overlapped with the texture for part-1,2,3 and 4 of the Bimba. (C) The un-compressed texture mapping result. (D)The compressed texture mapping result.\label{fig:bima}}
\end{figure*}

Figure \ref{fig:manface} and \ref{fig:bima} give a simple illustrations of the idea. Figure \ref{fig:manface}(A) show the surface mesh of a human face. Its texture mapping and texture image are shown in (B). Note that there is only one texture mapping in this case and that the texture image is of regular shape. The original textured surface is shown in (C). The proposed compression algorithm is applied, and the reconstructed textured surface after compression is shown in (D). Note that the compressed textured surface closely resembles to the original textured surface. Figure \ref{fig:bima}(A) shows the surface mesh of the bimba surface. Its texture mappings and texture images are as shown in (B). There are totally four texture mappings and texture images with regular shapes. The original textured surface is as shown in (C). We apply the proposed compression algorithm to this example. The reconstructed textured surface after the compression is as shown in (D). In both example, only 1\% of Fourier coefficients are saved. Again, the reconstructed textured surface closely resembles to the original one.

\begin{figure*}[t]
\centering
\includegraphics[height=3.7in]{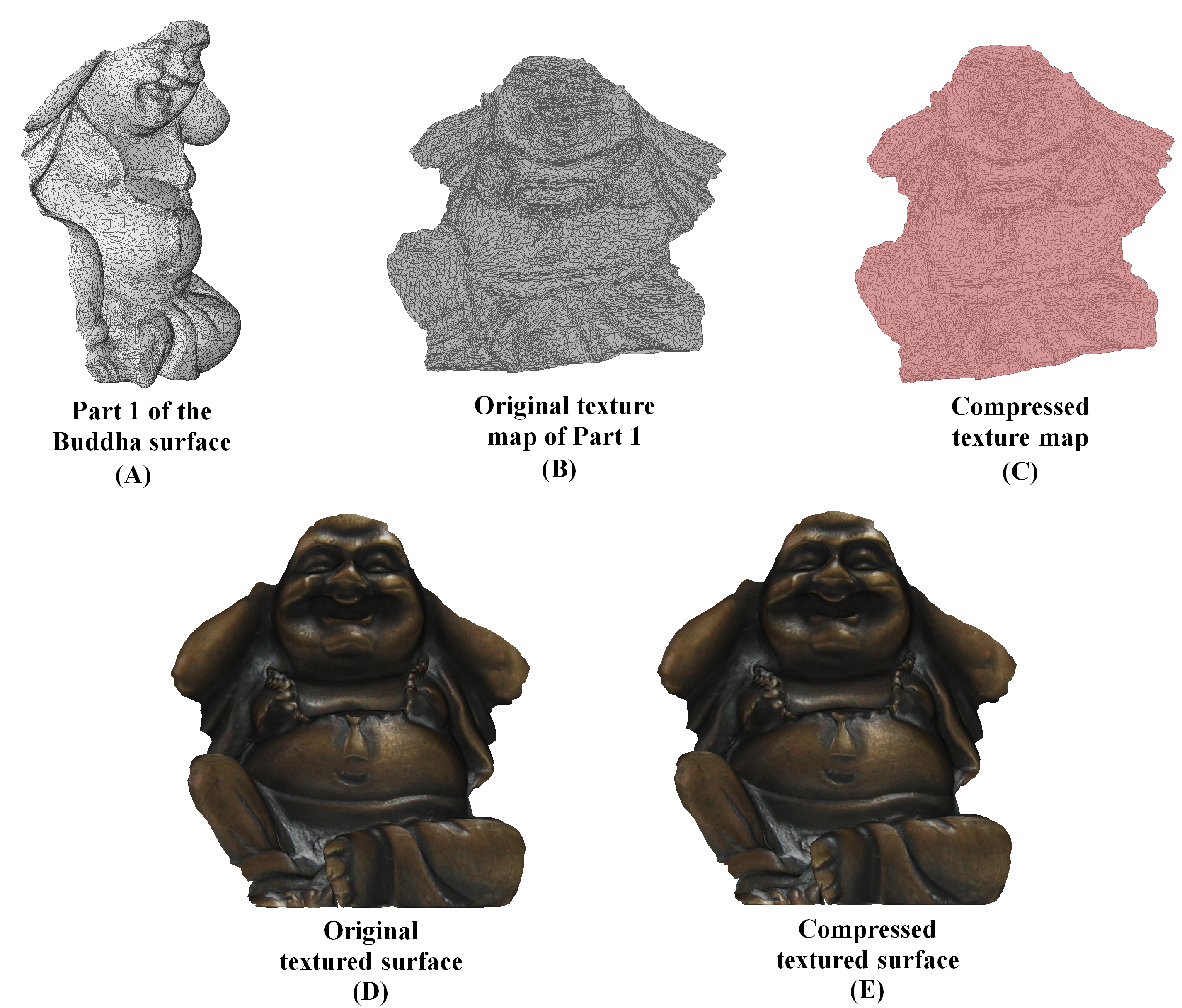}
\caption{Texture map compression for one part of Buddha. (A) shows the mesh of Buddha part 1. (B) shows the original texture map of part 1. (C) shows the compressed texture map of part 1 encoded by algorithm 5.1. (D) and (E) show the original textured surface and the one reconstructed from the compressed texture map in (C).$^2$\label{fig:buddhapatch}}
\end{figure*}

Figure \ref{fig:buddhapatch} shows a more complicated example whose texture image is of irregular shape, which is common in real situation. (A) shows one part of the Buddha surface mesh. It is being mapped to a texture image of irregular shape. The texture mapping is shown in (B). The compressed texture mapping using our proposed algorithm is shown in (C). (D) shows the original textured surface. (E) shows the textured surfaces reconstructed from the compressed texture mapping in (C). Note that there is no visual difference between the compressed textured surface and the original textured surface.  

\begin{figure*}[t]
\centering
\includegraphics[height=1.6in]{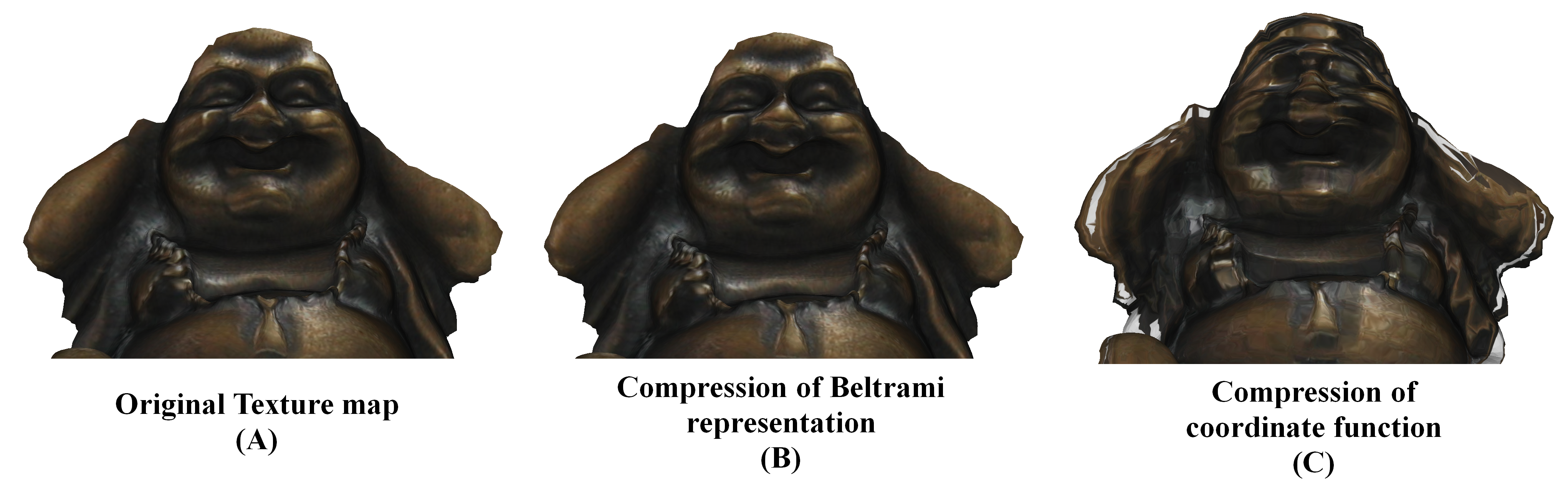}
\caption{Comparison between the results of compressing Beltrami representation and the coordinate function. (A) shows the original textured Buddha surface. (B) shows the Beltrami representation compression result. (C) shows the coordinate function compression result.\label{fig:buddha_compare}}
\end{figure*}

Note that compression of texture mappings with their coordinate functions generally does not work. Figure \ref{fig:buddha_compare}(A) shows the orginal textured buddha surface. (B) shows the compressed textured surface using the Fourier compression of the Beltrami representation. The reconstructed textured surface closely resemble to the original one. (C) shows the compressed texture surface using the Fourier compression of the coordinate functions, with the same compression ratio as in (B). Distortion of the texture can clearly be observed.  

\bigskip
\noindent Summary of the encoding and the decoding part of our proposed compression scheme are described as follow:

\noindent $\mathbf{Algorithm\ 5.1:}$ {\it( Encoding of the texture coordinates )}\\
\noindent $\mathbf{Input:}$ {\it Surface mesh: $M$; Texture coordinate function $f: M\to [0,1]\times[0,1]$}\\
\noindent $\mathbf{Output:}$ {\it Fourier coefficients $c_{j,k}$ in the truncated Fourier series; Coordinates of the boundary vertices of $T$.}\\
\vspace{-3mm}
\begin{enumerate}
\item {\it Obtain the harmonic parametrization, $\phi : M \rightarrow D$, by solving Equation \ref{eqt:harmonic2}}
\item {\it Compute the Beltrami representation $\mu_{f}$ of $f$ using Algorithm 4.1}
\item {\it Compress the Beltrami representation $\mu_f$ using Algorithm 4.3}
\item {\it Store the Fourier coefficients $c_{j,k}$ in the truncated Fourier series and coordinates of the boundary vertices of $T$.}
\end{enumerate}

\let\thefootnote\relax\footnotetext{\noindent 2. The texture mapping example is freely available on http://www.kunzhou.net/tex-models.htm}

\bigskip

\noindent $\mathbf{Algorithm\ 5.2:}$ {\it(Decoding of the texture coordinates)}\\
\noindent $\mathbf{Input:}$ {\it Surface mesh $M$; Fourier coefficients $c_{j,k}$ of the Beltrami representation; Coordinates of the boundary vertices}\\
\noindent $\mathbf{Output:}$ {\it Reconstructed texture coordinates of the surface mesh $M$.}\\
\vspace{-3mm}
\begin{enumerate}
\item {\it Apply inverse fast Fourier transform on $c_{j,k}$ to restore the Beltrami representation $\mu_{f}$ defined on triangular faces of $M$}
\item {\it Parametrize the surface mesh $M$ by the harmonic mapping $\phi$ to the unit square through solving Equation \ref{eqt:harmonic2}}
\item {Reconstruct the texture coordinates by using the Linear Beltrami solver (Equation \ref{eqt:linearB1} and \ref{eqt:linearB2}) subject to the boundary condition given by the texture coordinates of the boundary vertices.}
%\begin{scriptsize}
%\end{scriptsize}}
\end{enumerate}

{
\begin{table}
\label{table1}
\caption{Root mean square error}
\begin{tabular}{c||c|c|c|c|c|c}
&Vertices&Faces&RMSE(1\%)&CF(1\%)&RMSE(3\%)&CF(3\%)\\
\hline
Susan&5161&9999&$5.103e^{-4}$&12.29:1&$2.304e^{-4}$&8.32:1\\
Buddha-1&6523&12779&$3.009e^{-3}$&16.64:1&$1.731e^{-3}$&10.07:1\\
Buddha-2&1921&3609&$4.298e^{-3}$&7.19:1&$2.418e^{-3}$&5.67:1\\
Buddha-3&2385&4563&$3.551e^{-3}$&9.54:1&$1.870e^{-3}$&6.99:1\\
Buddha-4&2001&3842&$2.194e^{-3}$&10.21:1&$1.021e^{-3}$&7.33:1\\
Buddha-5&1985&3808&$2.766e^{-3}$&10.03:1&$1.551e^{-3}$&7.24:1\\
Zebra-1&7638&14683&$7.042^{-3}$&10.36:1&$4.898e^{-3}$&7.41:1\\
Zebra-2&1045&1924&$3.217e^{-3}$&5.71:1&$2.030e^{-3}$&4.73:1\\
Zebra-3&1287&2375&$3.914e^{-3}$&5.85:1&$2.311e^{-3}$&4.80:1\\
Zebra-4&7600&14622&$7.576e^{-3}$&10.53:1&$4.801e^{-3}$&7.50:1\\
Zebra-5&767&1414&$1.395e^{-3}$&5.82:1&$1.116e^{-3}$&4.79:1\\
\end{tabular}
\end{table}
}

\begin{figure*}[h!]
\centering
\includegraphics[height=4in]{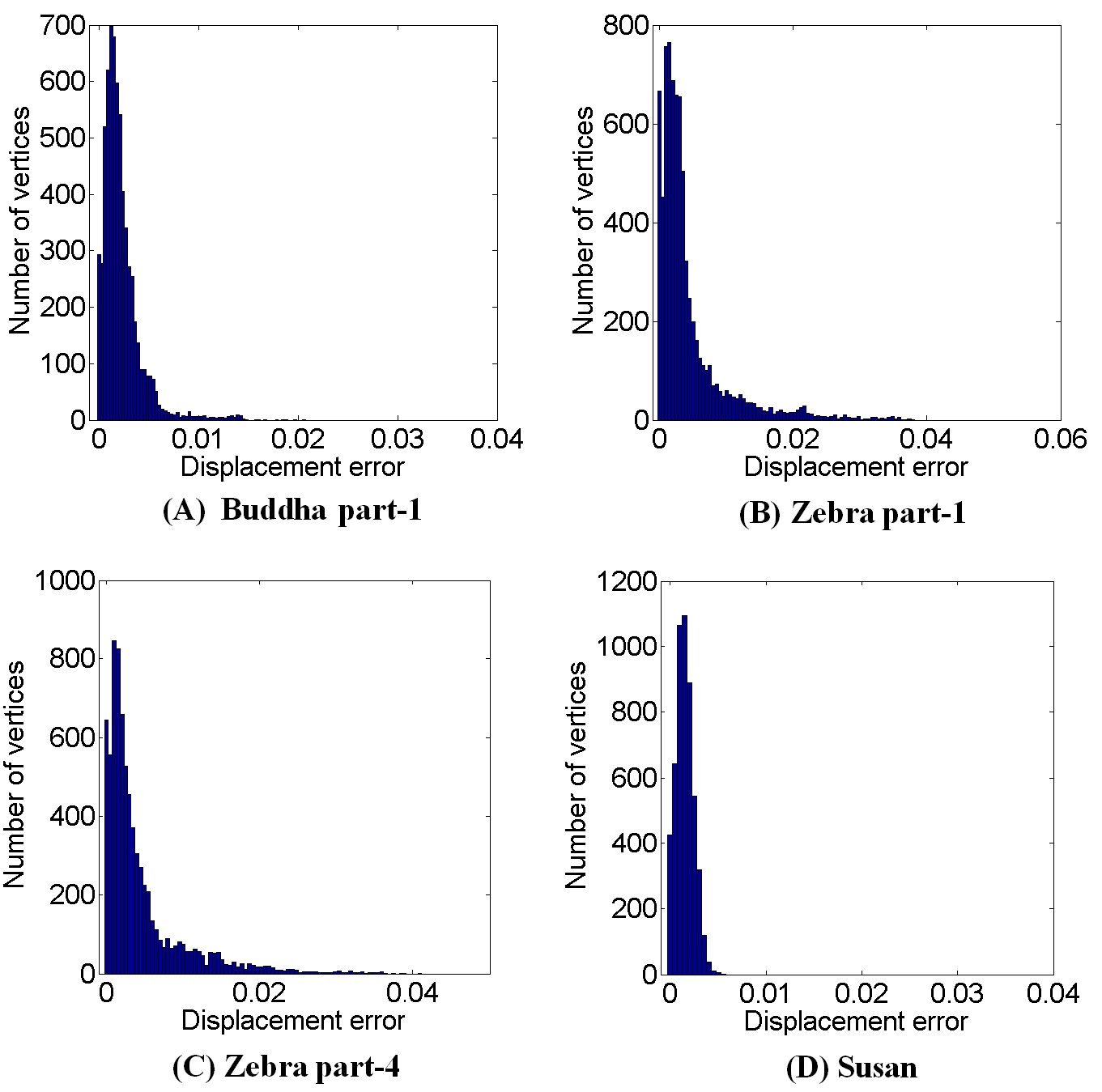}
\caption{Statistics of the displacement error of each vertices. (A)Buddha part-1. (B)Zebra part-1. (C)Zebra part-4. (D)Susan. The majority of the reconstructed texture coordinates has less than 1 relative percentage error. \label{fig:displacement_err_summary_renew}}
\end{figure*}

\begin{table}
\caption{Summary of the texture mapping compression result.}
\label{table2}
\begin{center}
\begin{tabular}{c||c|c|c}
&Susan&Buddha&Zebra\\
\hline
Number of vertices&5161&15138&20157\\
RMSE&$5.103e^{-4}$&$1.974e^{-3}$&$6.44e^{-3}$\\
Compression ratio&12.29:1&17.27:1&13.68:1\\
Data saving&91.86\%&94.21\%&92.69\%\\
Memory required&1.6406 kB&3.3691 kB &5.6289 kB\\
Original memory required&20.1548 kB&58.1883 kB&77.0027 kB\\
\end{tabular}
\end{center}
\end{table}

\begin{figure*}[h!]
\centering
\includegraphics[height=3.5in]{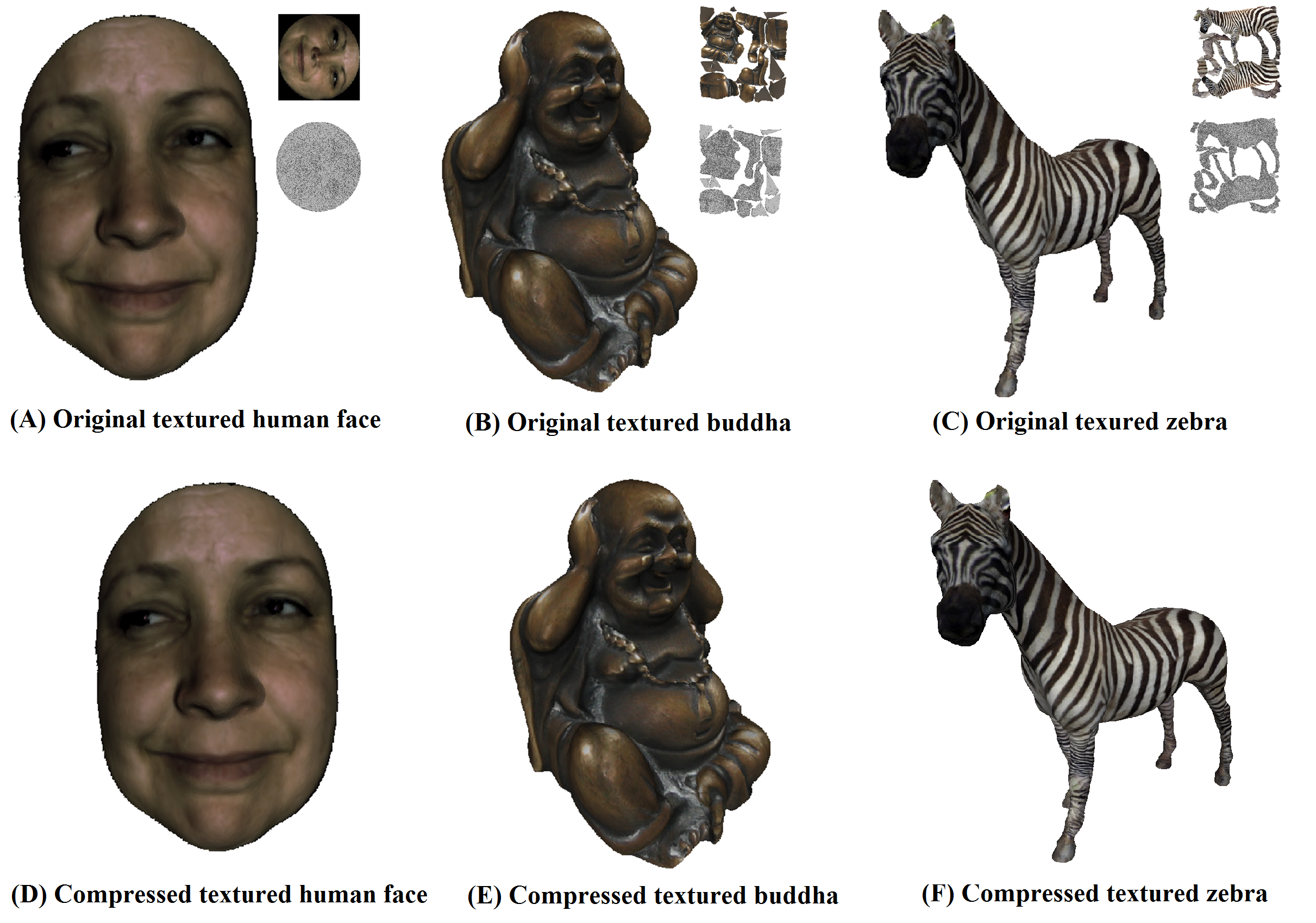}
\caption{Summary of compression results of Susan, Buddha and Zebra. The upper row shows the uncompressed textured surfaces with their corresponding texture maps and the texture images: (A)Susan (B)Buddha (C)Zebra. The lower row presents the compression results using 1\% Fourier coefficients: (D)Susan. (E)Buddha. (F)Zebra.\label{fig:summarytexture}}
\end{figure*}

Quantitative experiments have been carried out to demonstrate the performance of our proposed algorithm. Experiments are performed on three surface meshes, namely, 'Susan', 'Buddha' and 'Zebra' (See Figure \ref{fig:summarytexture}). 'Susan' is a simply-connected open surface with 5161 vertices. 'Buddha' and 'Zebra' are genus-0 closed surface with  15138 vertices and 20157 vertices respectively, which are partitioned into several simply-connected open surfaces. Since some partitions in 'Buddha' and 'Zebra' contain very few vertices, we applied the compression algorithm only on major parts with more vertices. We tested the proposed algorithm with 1\% and 3\% of the Fourier coefficients(FC).

To measure the accuracy of the compression scheme, we compute the root mean square error(RMSE), defined by 
\begin{equation}\label{RMSE}
RMSE = \sqrt{\frac{1}{n}\sum\limits_{i=1}^n \|f(v_i) - \tilde{f}(v_i) \|_{1}},
\end{equation}
\noindent where $n$ stands for the number of vertices, $f$ and $\tilde{f}$ are the original and the reconstructed texture map respectively. We also compute the compression ratio(CR) to quantify the compression efficiency of the algorithm. The compression results for each partition of the surface meshes are shown in Table \ref{table1}.  Results shows that the reconstructed textured surface after compression is very close to the original data, with very small RMSE. Even in the case when only 1\% of Fourier coefficients are saved, the RMSE still remains in the order of $10^{-3}$.  As expected, the RMSE is smaller when 3\% of Fourier coefficients are saved (see column 5). The compression ratios are shown in column 4 and 6, which illustrate our proposed algorithm can significantly reduce the storage memory for texture mappings. Note also that the table shows the out-performance of CR in 'Buddha-1', 'Zebra-1', 'Zebra-4' and 'Susan'. This is expected as texture coordinates of the boundary vertices contribute a relatively small amount of total storage of the fine meshes. Hence, our compression algorithm, which compresses the texture coordinates in the interior, gives better compression results. Figure \ref{fig:displacement_err_summary_renew} presents the displacement error of the each reconstructed texture coordinates measured in the supreme-norm. It is observed that majority of the texture coordinates can be reconstructed almost losslessly.

In Table \ref{table2}, we study the compression performance of the whole mesh for the three surfaces. For each surface, 1\% of Fourier coefficients are stored. The average RMSE is computed which is defined as:
\begin{equation}
RMSE_{avg} = \sum\limits_{k=1}^P \frac{n_k}{N}RMSE(k),
\end{equation}
where $P$ is the number of parts in surface mesh, $N$ is the total number of vertices, $n_k$ stands for the number of vertices in part $k$ of the partitioned mesh and $RMSE(k)$ is the corresponding $RMSE$ calculated by formula \ref{RMSE}. Note that the average RMSE is small in all 3 cases. Therefore, there is no noticeable difference between the compressed textured surface and the original one.

\begin{figure*}[t]
\centering
\includegraphics[height=1.35in]{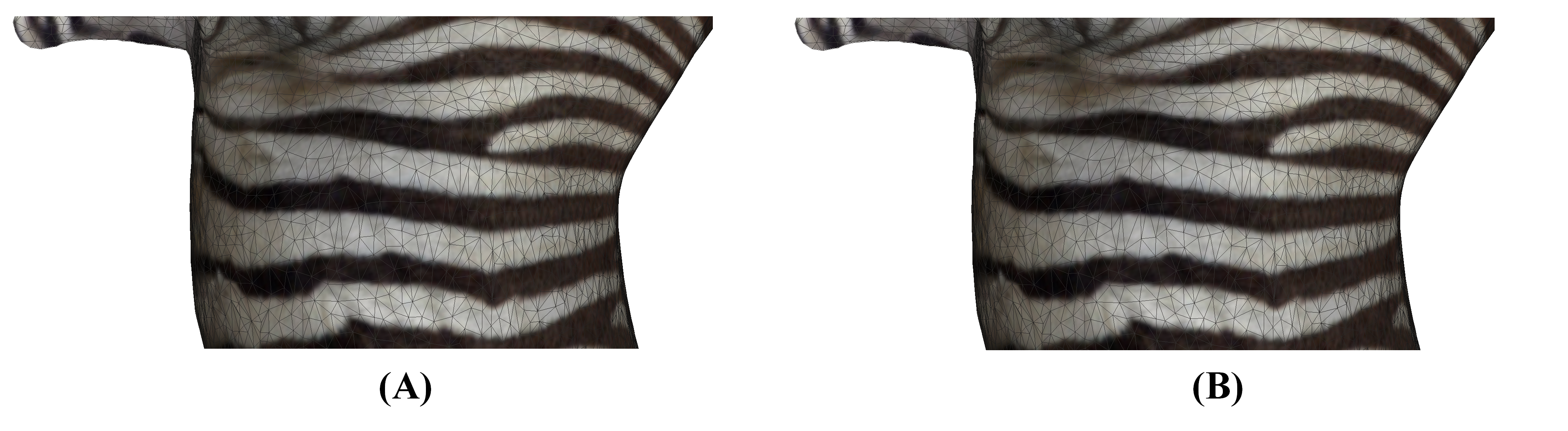}
\caption{A close look to Zebra part-1. (A)Uncompressed Texture mapping. (B)Compressed texture mapping with $4.9\times10^{-3}$RMSE.\label{CLOSE_LOOK}}
\end{figure*}

The actual storage memories needed are also listed. Note that after the compression, the storage memories of the texture surfaces are much less than that of the original data. The compression ratios are at least 12:1. This demonstrates the efficacy of our proposed method.

Finally, every smooth Beltrami coefficient corresponds to a smooth bijective quasi-conformal map. As a result, no abnormal texture (e.g. zaggy textures) would appear on the textured mesh. The diffeomorphic property is also a major reason why there is no noticeable difference between the original and the reconstructed texture mapping even with higher RMSE (See Figure \ref{CLOSE_LOOK}).

\subsection{Video Compression}
With the recent development of video compression techniques, real time digital television and Internet streaming video become practical. However, resolution of the video is limited by the channel bandwidth. Video compression is therefore an important field of research.

\begin{figure*}[t]
\centering
\includegraphics[height=1.4in]{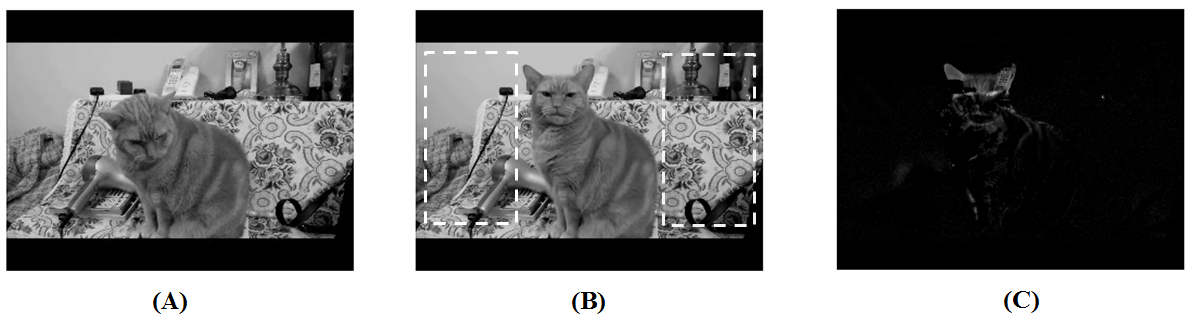}
\caption{Illustration of the redundant information appears in a video. (A) The $1^{st} frame$ (B) The $31^{st}$ frame. Area in the boxes are examples of duplicated information appear in the video.(C) The intensity difference between (A) and (B).\label{fig:paypay_with_words_fix}}
\end{figure*}

Different video compression algorithms have been proposed \cite{Hoang,LeGall,Wangvideo}. The basic idea of these algorithms is to remove temporal and spatial redundancies existing in a video sequence. For instance, the two images shown in Figure \ref{fig:paypay_with_words_fix}(A) and (B) are the $1^{st}$ and the $31^{st}$ frame of a video respectively. (C) shows the intensity difference between (A) and (B). The black regions correspond to the unchanged background. The unchanged background is an example of the temporary redundant information existing in the uncompressed video. In motion compensation techniques, instead of storing every image frames in the video, I-frame, P-frame and B-frame are introduced. Commonly, video frames are partitioned into frame sets called the Group of Picture(GOP):
$$I_1 B_1 B_2 P_1 B_3 B_4 P_2 B_5 B_6 P_3 B_7 B_8 P_4 B_9 B_{10}$$ I-frame is the reference frame, which is stored in the compressed image form. P-frame is called the predictive frame. It is encoded as a motion vector field(MV field) together with a residual. The MV field creates a prediction depicting how pixels in previous frame move. This prediction is then subtracted from the original frame to obtain the residual image. If the prediction is successful, the residual image can be represented with fewer bits than that of the original frame. B-frame is called the bi-directional predictive frame. It is encoded as the residual of the predicted image, which is obtained from the interpolated MV fields from previous and future frames.

On average, P-frames contributes 50\% less storage than that of I-frames. However, the required storage memory for the MV field is still significant when considering HD videos. For example, if a maximum of $4\times4$ block mode is used in a full HD $[1920\times1280]$ video, each P-frame requires $2.765\times10^6$ bits of memory to store the MV field. It thus calls for the need of compressing the MV field.

Suppose $F_1$ and $F_2$ are two image frames in a video sequence. The MV field $V:F_1\to \mathbb{R}^2$ is a vector field defined on $F_1$. It describes how pixels in $F_1$ should move to get the estimation of the image frame $F_2$. In particular, the MV field $V$ can be considered as a mapping $T:F_1\to F_2$ between $F_1$ and $F_2$ given by:

\begin{equation}
T(x,y) = (x,y) + V(x,y)
\end{equation}

By introducing a Delaunay triangulation to $F_1$, $T$ can be viewed as a piecewise linear homeomorphism between triangular meshes. Hence, every MV field $V$ can be represented by the Beltrami representation $\mu_T$ of $T$. Note that the harmonic parameterization of the meshes are not required in this case, since the triangular mesh is already a 2D rectangle embedded in $\mathbb{R}^2$. Using Algorithm 4.3, the MV field $V$ can be compressed by performing the Fourier compression of $\mu_T$. Storing the Fourier coefficients of the truncated Fourier series of $\mu_T$ requires much less storage memories than that of the MV field itself.

The reconstruction of the MV field from the Fourier coefficients $c_{j,k}$ is straightforward. Following the idea of Section 4.3, we first carry out the inverse fast Fourier transform to restore the Beltrami coefficient $\mu_T$. Using the Linear Beltrami solver as described in Algorithm 4.2, $T$ and hence the corresponding MV field $V$ can be reconstructed efficiently and accurately.

\bigskip

\noindent The encoding and decoding of the compression algorithm can be summarized as follows.

\noindent $\mathbf{Algorithm\ 5.3:}$ {\it (Encoding of the P-frame)}\\
\noindent $\mathbf{Input:}$ {\it Frame $F_1$ (Reference) and $F_2$ (P-frame)}\\
\noindent $\mathbf{Output:}$ {\it Coefficients $c_{j,k}$ in the truncated Fourier series}\\
\vspace{-3mm}
\begin{enumerate}
\item {\it Obtain the motion vector field in regular grid}
\item {\it Compute the Beltrami representation by Algorithm 4.1}
\item {\it Use Algorithm 4.3 to compress the Beltrami representation $\mu$ and store the coefficients $c_{j,k}$ in the truncated Fourier series.}
\end{enumerate}

\bigskip

\begin{figure*}[p]
\centering
\includegraphics[height=7.45in]{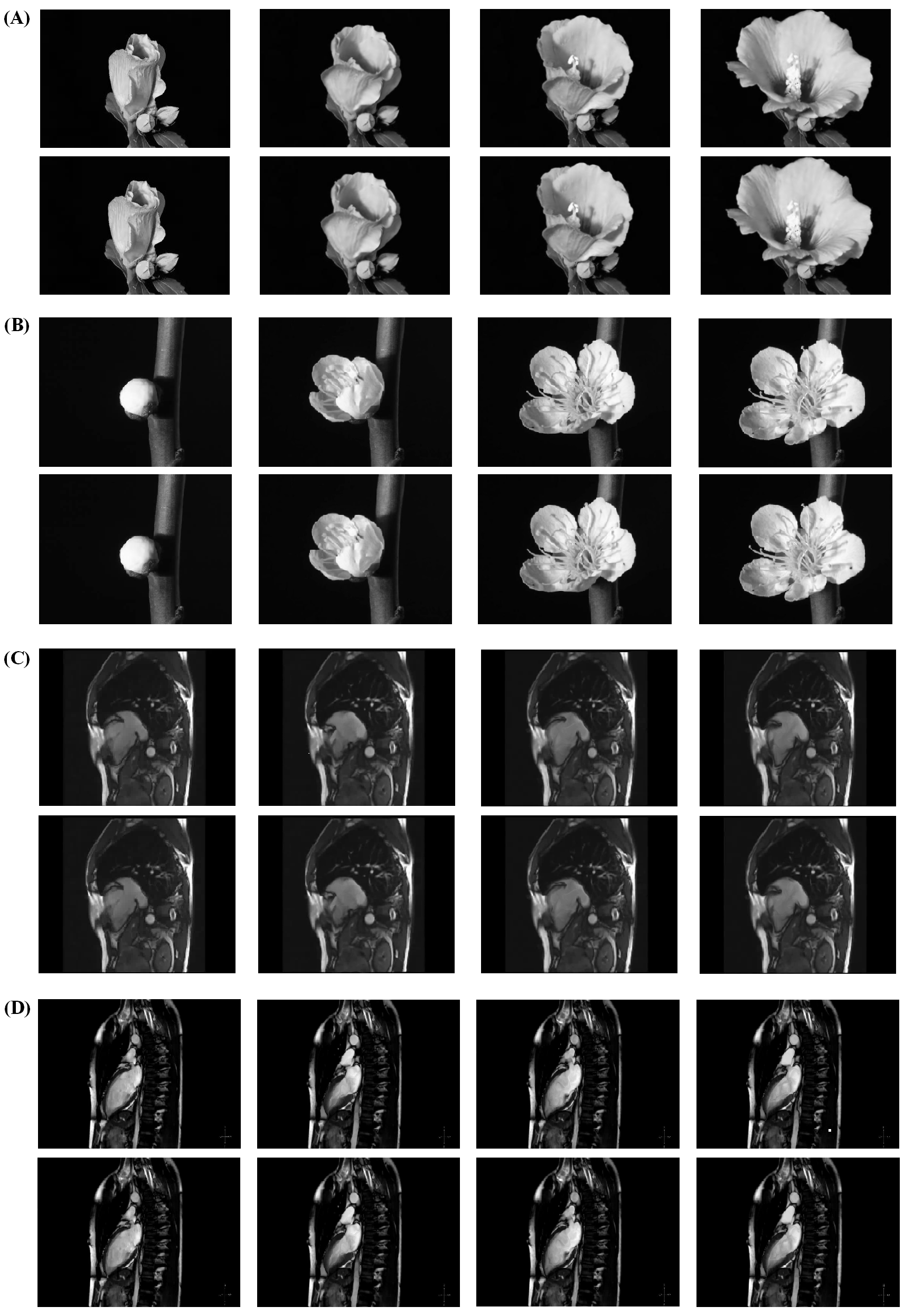}
\caption{Comparison between the uncompressed and the compressed video using 0.5\% of Fourier coefficients: (A)Flower 1. (B)Flower 2. (C)Heart 1. (D)Heart 2. Upper rows are some original $4^{th}$ P-frames in each frame set. The lower rows show the corresponding compressed P-frames.$^3$\label{fig:summary_4video}}
\end{figure*}

\begin{figure*}[t]
\centering
\includegraphics[height=1.45in]{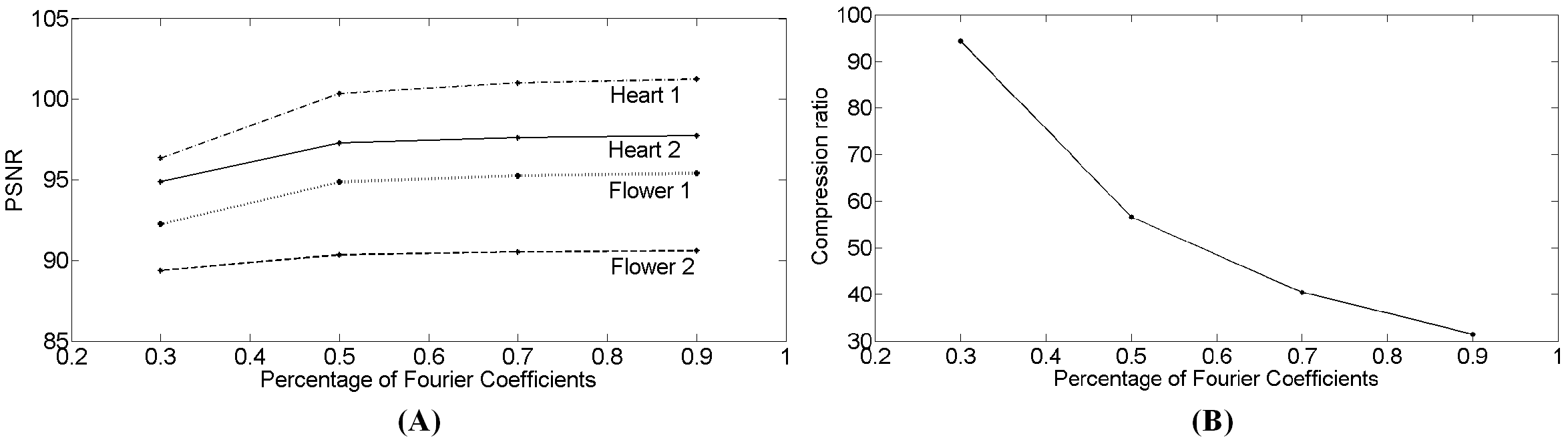}
\caption{(A)The average PSNR between the uncompressed and the compressed P-frames for the four videos with difference percentage of Fourier coefficients saved. (B)The corresponding compression ratio obtained.\label{fig:cr_psnr}}
\end{figure*}

\begin{figure*}[h!]
\centering
\includegraphics[height=3in]{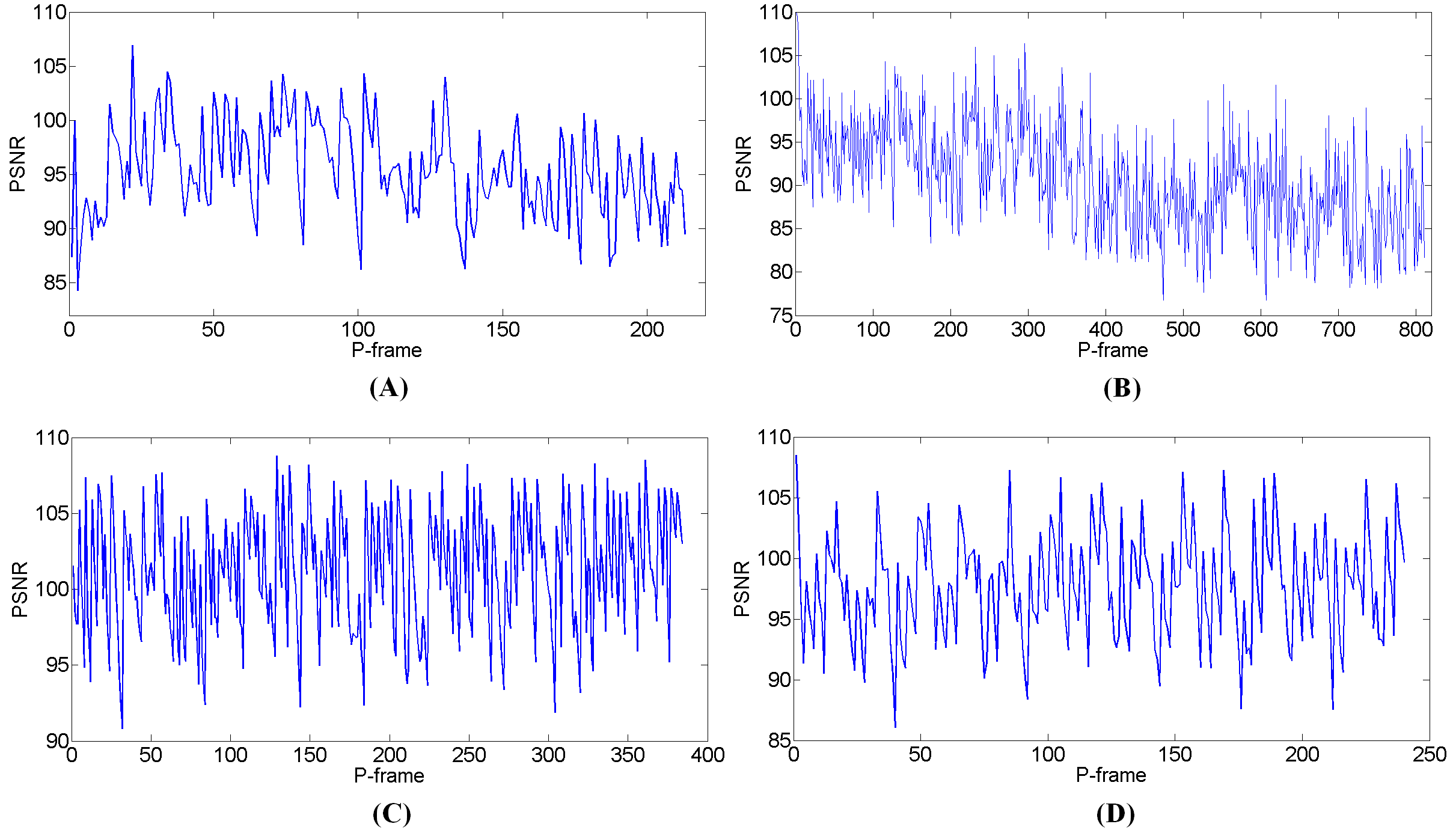}
\caption{PSNR between each uncompressed and compressed P-frames. (A)Flower 1. (B)Flower 2. (C)Heart 1. (D)Heart 2. \label{fig:psnr_vs_frame}}
\end{figure*}

\noindent $\mathbf{Algorithm\ 5.4:}$ {\it (Decoding of the P-frame)}\\
\noindent $\mathbf{Input:}$ {\it Frame $F_1$; Coefficients $c_{j,k}$ in the truncated Fourier series}\\
\noindent $\mathbf{Output:}$ {\it Motion vector field $V$}\\
\vspace{-3mm}
\begin{enumerate}
\item {\it Apply inverse fast Fourier transform to restore the Beltrami coefficient $\tilde{\mu}$}
\item {\it Perform Linear Beltrami solver as described in Algorithm 4.2 to obtain $(\tilde{x},\tilde{y})$}
\item {\it Compute $V$ by $V(x,y) = (\tilde{x},\tilde{y}) - (x,y)$}
\end{enumerate}

\bigskip

To study the performance of our proposed algorithm, experiments have been carried out on real videos. In our experiments, all B-frames in the GOP are taken away for simplicity. Instead, we consider the following frame set:
$$ I_1 P_1 P_2 P_3 P_4$$
MV fields in each P-frames are obtained from the previous I-frame or P-frame, which are then compressed by our proposed algorithm. In order to study the performance of our algorithm without adding any residual, per-pixel MV field is used. It should be noted that the proposed algorithm can also be applied to block-based MV fields in exactly the same way. However, residual has to be added. In the decoding process, errors may be introduced to the MV fields after the compression. As a result, the reconstructed P-frame $\widetilde{P_i}$ might not be identical to the original frame $P_i$. Meanwhile, the decoding of $P_{i+1}$ is based on the reconstructed P-frame $\widetilde{P_i}$ rather than the original $P_i$. The reconstruction error would be accumulated. However, experimental results show that the accumulated errors are small that the effect to the overall results would be negligible. 

\begin{figure*}[t]
\centering
\includegraphics[height=3in]{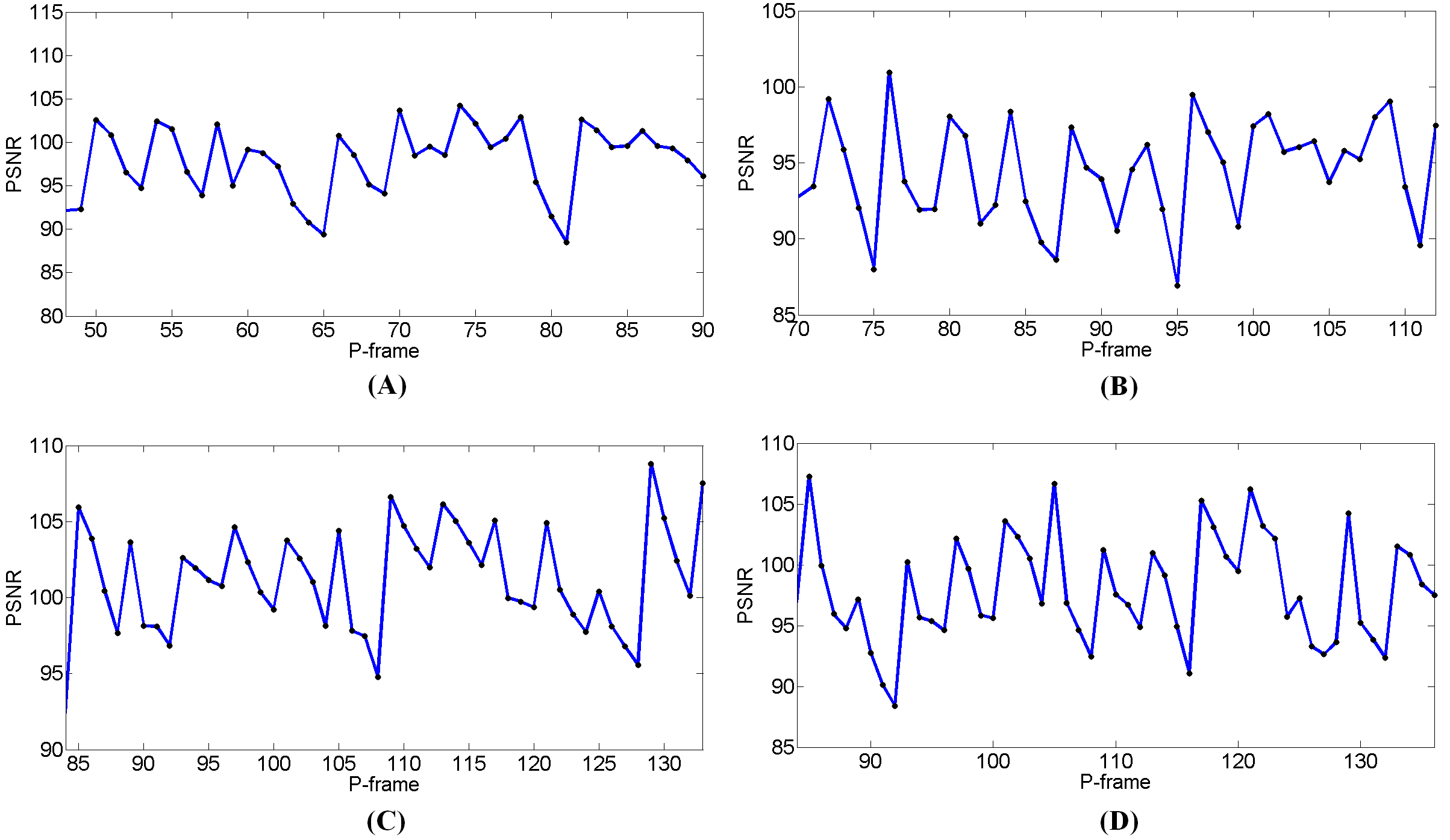}
\caption{A closer look to the plots in Figure \ref{fig:psnr_vs_frame}. Patterns of decreasing PSNR with a period of four frames can be observed. This is the consequence of the accumulation of errors from previous P-frame reconstruction. (A)Flower1. (B)Flower2. (C)Heart1. (D)Heart2.\label{fig:closelook_summary}}
\end{figure*}

Experiments have been carried out on four video clips, namely 'Flower 1', 'Flower 2', 'Heart 1' and 'Heart 2'. Their resolutions are $[360\times262]$, $[512\times384]$, $[600\times480]$ and $[1280\times720]$ respectively. Different percentages of Fourier coefficients have been used. Figure \ref{fig:summary_4video} shows some original P-frames of the 4 videos and their corresponding compressed P-frames using our proposed algorithm. Here, 0.5\% of Fourier coefficients are used. As shown in the figure, no visible differences can be observed after compression. It means, it demonstrates the effectiveness and accuracy of our proposed algorithm in compressing MV field.

We also examine the performance of our algorithm quantitatively. Figure 5.14 shows the the peak signal-to-noise ratio (PSNR) between the reconstructed P-frames and the original P-frames. The PSNR is defined by
\begin{equation}\label{eqt:PSNR}
PSNR = 10 \log_{10}\frac{255^2NM}{\sqrt{MSE}},
\end{equation}
with
\begin{equation}\label{eqt:MSE}
MSE = \frac{1}{MN}\sum\limits_{i=1}^M\sum\limits_{j=1}^N\left[\widetilde{P}(i,j)-P(i,j)\right]^2,
\end{equation}
where the size of the frames is $M \times N$. As shown in Figure \ref{fig:cr_psnr}(A), the average PSNR in all four cases are higher than 90 when 0.5\% Fourier coefficients are captured. In other words, only a negligible increase in residual will be induced by the error of the reconstructed MV field. As a result, the required storage memory for the residual image will not be much affected after compression with our algorithm.

Figure \ref{fig:cr_psnr}(B) presents the corresponding compression ratio for different percentage of Fourier coefficients stored. When 0.5\% coefficients are saved, compression ratio of the MV field can be as high as $56 : 1$. The high compression ratio of the MV field can effectively improve the compression ratio of the existing video compression algorithm such has MPEG and H.264. After the compression, the storage memory of the MV field is almost negligible compare with the total file size. For example, suppose the MV fields contribute $1/3$ total storage of the compressed video. Using our algorithm to compress the MV field, the storage requirement of the compressed video can be reduced by around 32\%.

Figure \ref{fig:psnr_vs_frame} shows the PSNR between each pair of the compressed and original P-frame. It is observed that PSNR stays close to about 90. It means the compressed P-frames closely resemble to the original ones. Figure \ref{fig:closelook_summary} shows the zoom-in of the plots in Figure \ref{fig:psnr_vs_frame}. Decreases in the PSNR can be observed in each frame set. This is expected since the errors in MV fields are accumulated in each cycle. However, the decreases are very small. It means the effect of the accumulated errors in the MV fields to the overall result is negligible.

\section{Conclusion}
We address the problem of compressing surface homeomorphisms, which has important applications in computer graphics and imaging. Surface homeomorphisms are usually represented and stored by their 3D coordinate functions. It often requires lots of storage memory, which causes inconvenience in data transmission and data storage. In this paper, we propose an effective algorithm for compressing piecewise linear bijective surface maps between meshes using their Beltrami representations. The Beltrami representation is a complex-valued function defined on triangular faces of the surface mesh with supreme norm strictly less than 1. Under suitable normalization, there is a 1-1 correspondence between the set of surface homeomorphisms and the set of Beltrami representations. Given a Beltrami representation, the associated bijective surface map can be exactly reconstructed using the Linear Beltrami Solver introduced in this paper. Since the Beltrami representation has very few constraints, it can be easily combined with the Fourier approximation to compress bijective surface map without distorting the bijectivity of the map. This significantly reduces the storage requirement for surface maps. In this paper, we proposed to apply the algorithm to texture map compression and video compression. With our algorithm, the storage requirement for textured surfaces can be significantly reduced, while well preserving the quality of the original data. Our algorithm can also be applied to compressing motion vector fields for video compression. After compressing the motion vector field, the compression ratio of the state-of-the-art video compression algorithms can be significantly improved. Experiments on real textured surfaces and videos demonstrate the effectiveness and efficacy of our proposed algorithms. 

\section*{Acknowledgments}
Lok Ming Lui is supported by RGC GRF (Porject ID: 2130271)and CUHK Direct Grant (Project ID: 2060413).

\end{document}